\newif\ifTwoColumn%
\newif\ifSUBMIT%
\newif\ifCOMMENTS%
\newif\ifFIGs%
\newif\ifFIGoneColumn%
\let\ifSUBMIT\iftrue%
\let\ifCOMMENTS\iftrue%
\let\ifFIGoneColumn\iftrue%

\documentclass[journal=jcisd8,manuscript=article]{achemso} %JCIM
\setkeys{acs}{articletitle = true}
\mciteErrorOnUnknownfalse
\setkeys{acs}{etalmode = truncate, maxauthors = 10}

\usepackage{amsmath,graphicx}
\SectionNumbersOn

\usepackage{graphicx}
\usepackage{amsmath}
\usepackage{xcolor}
\usepackage{todonotes}
\usepackage{menukeys}
\usepackage{hyperref}
\setkeys{acs}{doi = true}
\newcommand{\doi}[1]{\href{https://doi.org/#1}{\nolinkurl{#1}}}
\usepackage{acronym}
\usepackage{array,mathtools,amssymb,booktabs}
\usepackage{physics}
\usepackage{acronym}
\usepackage{placeins}

\ifSUBMIT%
  \ifCOMMENTS%
    \usepackage{color}    %%% NOTE this is necessary in pdflatex
    \usepackage[normalem]{ulem}
    
    \def\STRIKE#1{{\color{red}\sout{#1}}}
    
    \def\STRIKE#1{}
    
    \def\NSTRIKE#1{{\color{red}\sout{#1}}}
  \else
    
    \def\STRIKE#1{}
    
    \def\NSTRIKE#1{}
  \fi
\else
  \usepackage{color}    %%% NOTE this is necessary in pdflatex
  \usepackage[normalem]{ulem}
 \definecolor{mygreen}{RGB}{0,180,0}    %%% NOTE that this isn't necessary.
  
  \def\STRIKE#1{{\color{red}\sout{#1}}}
  
  \def\STRIKE#1{\relax}
  
  \def\NSTRIKE#1{{\color{blue}\sout{#1}}}
\fi

\NewDocumentCommand\eg{}{e.\,g.}

\title[CNN Classification of DNA Origami]%
{DNA Origami Nanostructures Observed in Transmission Electron Microscopy Images can be 
Characterized through 
Convolutional Neural Networks}
\author{Xingfei Wei}
\affiliation{Department of Chemistry, 
Johns Hopkins University, Baltimore, Maryland 21218, USA}
\author{Qiankun Mo}
\affiliation{Department of Chemistry, 
Johns Hopkins University, Baltimore, Maryland 21218, USA}

\author{Chi Chen}
\affiliation{Department of Biological Engineering, Massachusetts Institute of Technology, 
Cambridge, Massachusetts 02139, United States}
 \author{Mark Bathe}
\affiliation{Department of Biological Engineering, Massachusetts Institute of Technology, 
Cambridge, Massachusetts 02139, United States}
 
\author{Rigoberto Hernandez}
\email{r.hernandez@jhu.edu}
\affiliation{Department of Chemistry, 
Johns Hopkins University, Baltimore, Maryland 21218, USA}
\alsoaffiliation{Department of Chemical \& Biomolecular Engineering, 
Johns Hopkins University, Baltimore, Maryland 21218, USA}
\alsoaffiliation{Department of Materials Science and Engineering, 
Johns Hopkins University, Baltimore, Maryland 21218, USA}


\date{\today}
\begin{document}

%%%%%%%%%%%%%%%%%%%%%%%%%%%%%%%%%%%%%%%%%%%%%%%%%%%%%%%%%%
\newlength\figurewide
\ifFIGoneColumn
  \figurewide=.5\columnwidth
\else
  \figurewide=.9\columnwidth
\fi

%%%%%%%
\acrodef{ACCESS}{Advanced Cyberinfrastructure Coordination Ecosystem: Services \& Support}
\acrodef{NP}{nanoparticle}
\acrodef{QD}{quantum dot}
\acrodef{ssDNA}{single-stranded DNA}
\acrodef{ML}{machine learning}
\acrodef{TL}{transfer learning}
\acrodef{AI}{artificial intelligence}
\acrodef{FC}{fully-connected layer}
\acrodef{CL}{convolutional layer}
\acrodef{CNN}{convolutional neural network}
\acrodef{MD}{molecular dynamics}
\acrodef{CG}{coarse-grained}
\acrodef{DT}{decision trees}
\acrodef{RF}{random forest}
\acrodef{MC}{Monte Carlo}
\acrodef{LAMMPS}{Large-scale Atomic Molecular Massively Parallel Simulator}
\acrodef{LJ}{Lennard-Jones}
\acrodef{WCA}{Weeks-Chandler-Andersen}
\acrodef{SI}{Supporting Information}
\acrodef{SAv}{streptavidin}
\acrodef{TEM}{transmission electron microscopy}

%%%%%%
\begin{tocentry} %3.25 inches by 1.75 inches (approx. 8.25 cm by 4.45 cm).  TOC ACS format
\ \\
\includegraphics[clip=true,clip=true]{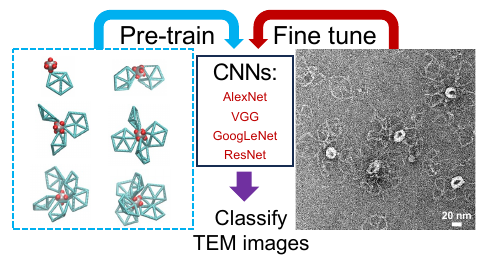}
\ \\
\end{tocentry}
%%%%%%

\begin{abstract} 

\Ac{AI} models remain an emerging strategy
to accelerate materials design and development.
We demonstrate that \ac{CNN} models can characterize
DNA origami nanostructures employed in programmable self-assembling,
which is important in many applications such as in biomedicine.
Specifically, we benchmark the performance of 9 \ac{CNN} models---%
{\it viz.}
AlexNet, GoogLeNet, VGG16, VGG19, ResNet18, 
ResNet34, ResNet50, ResNet101, and ResNet152%
---to characterize the ligation number of DNA origami nanostructures in \ac{TEM} images.
We first pre-train \ac{CNN} models using a large image dataset 
of 720 images from our \ac{CG} \ac{MD} simulations.
Then, we fine-tune the pre-trained \ac{CNN} models,
using a small experimental \ac{TEM} dataset with 146 \ac{TEM} images.
All \ac{CNN} models were found to have
similar computational time requirements,
while their model sizes and performances are different.
We use 20 test \ac{MD} images to demonstrate that 
among all of the pre-trained \ac{CNN} models
ResNet50 and VGG16 have the highest and second highest accuracies.
Among the fine-tuned models,
VGG16 was found to have the highest agreement on the
test \ac{TEM} images.
Thus, we conclude that fine-tuned VGG16 models can quickly
characterize the ligation number of nanostructures in large \ac{TEM} images.

\end{abstract}

%keywords: DNA nanotechnology, DNA origami, TEM image, neural network, transfer learning 

\acresetall

\section{Introduction}

Self-assembled nanostructures using DNA-based materials have emergent applications in
biomedicine \cite{CFan2019rev,Engelen2021,Bathe2023} 
and computing materials.\cite{CFan2015,hern21b,CFan2021rev}
DNA nanotechnology enables materials self-assembly at nanoscale accuracy
using engineered DNA-based building blocks. \cite{seeman1982, Bathe2016, Bathe2017}
A designed \ac{ssDNA} chain with a unique DNA sequence
can be folded into a pre-selected 2D or 3D nanostructure---%
DNA origami.\cite{rothemund2006,douglas2009,Bathe2016,Bathe2017}
The structural stability of planar 2D DNA origamis is crucial for 
developing functional materials, 
e.g. fabricating 2D arrays of \acp{QD}. \cite{bathe2022wang,bathe2023cc}
Using the dehydration and rehydration process, 
Chen et al.\cite{bathe2023cc}
have achieved ultrafast self-assembly of 2D regular arrays of \acp{QD}  
binding to rectangular DNA origamis on a surface at large-scale.
Recently, we have demonstrated that 
by engineering the positions and numbers of biotin binding sites on the DNA origami,
we can control the self-assembled 3D hierarchical nanostructures of 
DNA origamis and \acp{QD}.\cite{hern22j,hern24g}
However, characterizing nanostructures in \ac{TEM} images
according to the number of 
DNA origamis attached to a \ac{QD}---which
we call their ligation numbers---%
with reasonable consistency and accuracy
is a big challenge.

In the computer vision research area, 
\acp{CNN}---{\it viz.} AlexNet \cite{alexnet12},
GoogLeNet \cite{googlenet15},
VGG \cite{vgg14},
and ResNet \cite{resnet15}---%
have been extensively used to characterize objects in images.
Often standard open datasets of images containing 
objects from our daily life
are used to train and compare different \acp{CNN},
e.g. ILSVRC \cite{alexnet12,russakovsky15}, 
Caltech-101/256 \cite{caltech101,caltech256}, 
and CIFAR-10/100.\cite{cifar10}
In 2012, the AlexNet architecture was first published,
achieving the lowest error rate in the ILSVRC-2012 competition 
with only 5 \acp{CL} and 3 \acp{FC}.\cite{alexnet12}
In 2014, the VGG architecture,\cite{vgg14}
which uses 13 to 16 \acp{CL} and 3 \acp{FC},
significantly improved the classification accuracy.
However, 
it was found that for much deeper neural networks, 
the training and testing errors can increase 
with increasing the number of layers.\cite{resnet15} 
Using the inception method,
after carefully tuning the depth and width of the network
up to 21 \acp{CL} and 1 \ac{FC},
GoogLeNet managed to improve the accuracy even further.\cite{googlenet15}
In 2015, the ResNet architecture was developed to solve much deeper neural networks.
It uses shortcuts to jump over layers, but it also increased the
model size to 151 \acp{CL} and 1 \ac{FC}.\cite{resnet15} 

Previously, nanostructures of 
containing
monomeric, dimeric, trimeric, tetrameric, pentameric and, hexameric
DNA origamis
were self-assembled 
using biotinylated pentagonal pyramid wireframe DNA origamis
and \ac{SAv} functionalized \acp{QD}.\cite{hern22j,hern24g} 
This self-assembly process
has also been confirmed through 
\ac{CG} \ac{MD} simulations.\cite{hern24g}
We found that by engineering the biotin binding sites on the DNA origami
we can control the structure of the self-assembled nanostructures.\cite{hern24g}
Although the nanostructures can be directly characterized in our model,
characterization of these nanostructures in \ac{TEM} images can
suffer from inefficiency, inconsistency or inaccuracy.
Recent work by many 
groups\cite{jlee2021,nikishin2021,koyama2021,sytwu2022,ywang2023,lliu2023,gumbiowski2024}
has shown that \ac{AI} models---such as deep \acp{CNN}---can 
characterize experimental data with the benefits of
being fast and avoid observer bias
while being generally available through open-access.
For example, 
\acp{CNN} were used for characterizing
structures,\cite{sytwu2022}
surface dispersion locations,\cite{koyama2021}
3D atomic structures,\cite{jlee2021}
and crystallinity \cite{gumbiowski2024}
of \acp{NP}.
\acp{CNN} have also been used in \ac{TEM} image analysis to
trace the source origins of magnetic \acp{NP} in atmosphere PM$_{2.5}$,\cite{lliu2023} 
characterize DNA origami conformations,\cite{ywang2023}
and detect the morphology of extracellular vesicles.\cite{nikishin2021} 
Our recent work has demonstrated applications of \ac{ML} and \ac{AI} models in
nanoparticle-biological systems \cite{hern22m,hern24e} 
and developing energetic materials. \cite{hern24b,hern25a}

In this work,
we developed a general workflow 
for integrating simulation data with experimental data
to train \acp{CNN} by a \ac{TL} method to 
characterize nanostructures in \ac{TEM} images
according to their ligation numbers---namely, the number of origamis
attached to the \ac{QD}.
We pre-trained 4 types of benchmarking \acp{CNN}---{\it viz.}
AlexNet, \cite{alexnet12}
GoogLeNet, \cite{googlenet15}
VGG, \cite{vgg14}
and ResNet\cite{resnet15}---%
using nanostructure images of DNA origamis and \acp{QD} from \ac{MD} simulation trajectories.
We then fine-tuned the \acp{CNN} to characterize nanostructures in \ac{TEM} images
using an experimental \ac{TEM} image dataset.
Across several different \acp{CNN},
we report the architecture, parameter size, training time cost, 
and the accuracies for training, validation and testing.

\section{Methods}

\begin{figure}
\centering
\includegraphics[clip=true,scale=0.25,width=1\linewidth]{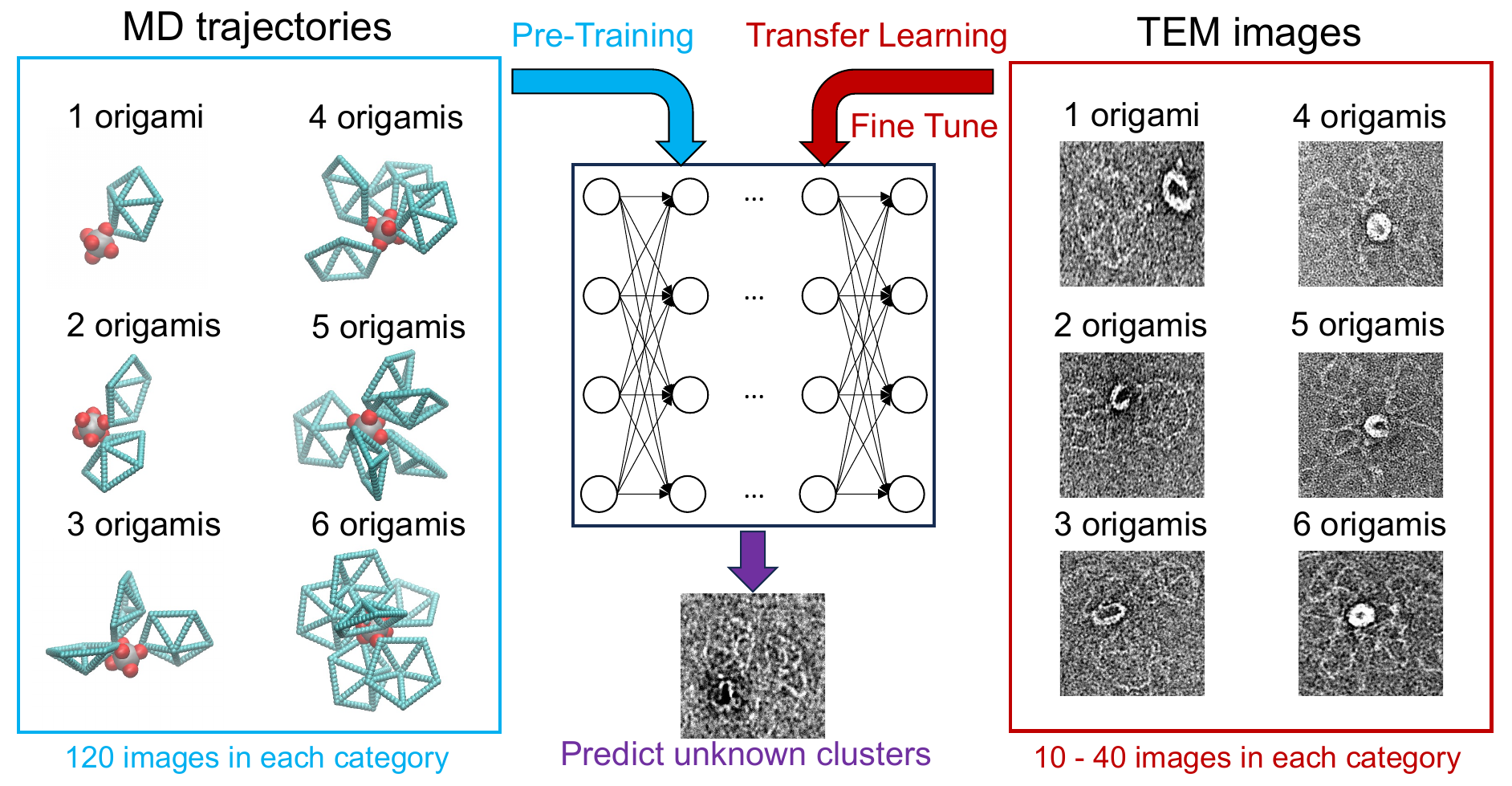}
\caption{
Scheme for predicting unknown nanostructures from a TEM image:
at left, the \ac{AI} models are pre-trained
using a large number of \ac{CG} \ac{MD} simulation images;
at right, the models are fine-tuned by the \ac{TL} method
using a small amount of experimental \ac{TEM} images;
and 
at center, unknown nanostructures in the \ac{TEM} images
are predicted using the fine-tuned models.
In the \ac{CG} \ac{MD} model on the left, \cite{hern24g} 
the DNA origamis are in green, biotin binding sites are in blue,
the \ac{QD} is in gray, and \ac{SAv} is in red.
The biotin bind sites are not visible 
because they are embedded into \ac{SAv}.
}
\label{fig:scheme}
\end{figure}

\subsection{Simulation and experimental datasets}

Details about
the \ac{CG} \ac{MD} simulation model  and experimental methods for 
self-assembling pentagonal pyramid DNA origamis and \acp{QD}
employed here can be found in previous work.\cite{hern22e,hern22j,hern24g}
We demonstrate the classification approach introduced here
using nanostructures self-assembled from
a collection of DNA origamis---%
{\it viz.} monomer, dimer, trimer, tetramer, pentamer, and hexamer---%
functionalized at the outer vertex 
of a single biotin binding site 
and binding to \ac{SAv} capped \acp{QD}.
We prepared a dataset 
containing images of nanostructures
from both simulations and experiments
whose ligation numbers varied from 1 to 6;
see Figure~\ref{fig:scheme}.
All the \ac{CG} \ac{MD} simulations were propagated
using the \ac{LAMMPS} package.\cite{plimpton95}
We used the \ac{WCA} potential to simulate the interactions between
\ac{CG} particles in DNA origamis and \acp{QD}.\cite{hern24g,wca1971}
The binding interaction between \ac{SAv} and biotin
was described by the \ac{LJ} potential.\cite{hern24g}
In simulation, the self-assembly process 
approached equilibrium
after a typical simulation time of 1500 $\mu$s.
The resulting nanostructures were visualized and saved using VMD software.\cite{vmd}
This led to the construction of
a dataset consisting of 120 \ac{MD} images---%
at $\sim500\times500$ pixels in size using the VMD render tool.
It was split into training and validation
subsets in a ratio of 80:20.

In experiment, 
the pentagonal pyramid DNA origamis were folded in a solution of 
12 mM MgCl$_2$, 1$\times$TAE buffer, 15 nM pF1A scaffold, 
and 150 nM staples.\cite{hern24g,shepherd2019}
DNA origamis were subsequently annealed on a Bio-Rad T100 thermocyler for 2 hr
and purified using an ultracentrifugal Amicon 100 kDa filter. 
The \ac{SAv} capped \acp{QD} were purchased from Thermo Fisher Scientific. 
DNA origamis and \acp{QD} were incubated together at room temperature overnight
to make self-assembled nanostructures through the biotin-\ac{SAv} interactions.
Large \ac{TEM} images with many nanostructures were taken using 
ThermoFisher FEI Tacnai Spirit \ac{TEM} at 120 kV.
Small images in PNG format with a single nanostructure 
were chopped from the large \ac{TEM} images
using the GIMP software.\cite{gimp}
Due to the limited amount of experimental data,
we chopped 10 to 40 \ac{TEM} images for each class 
at $\sim500\times500$ pixels in size
to fine-tune the pre-trained \acp{CNN}.

In general, hundreds of \ac{TEM} images are
required for training, validation, and testing \acp{CNN}.
See for example, Nikishin et. al. \cite{nikishin2021} 
which used 
138 (training) + 25 (validation) + 25 (testing) vesicle images
and Wang et. al. \cite{ywang2023} which used 
644 (training) + 4728 (validation) + 1000 (testing) particle images.
As has been well documented,
the Caltech-101 dataset has 101 categories with 50 images in most categories
and the size of each image is $\sim300\times200$ pixels.\cite{caltech101}
Meanwhile, the Caltech-256 dataset has 256 categories with more than 
80 images per category.\cite{caltech256}
In comparison to these two open datasets,
the number of images we obtained from \ac{MD} simulation for each category
is larger and hence should be large enough for training \acp{CNN}.
However, 
the number of our \ac{TEM} images is not large enough for directly training \acp{CNN},
and its collection was limited by the difficulty of 
measuring and preparing the experimental data.

We developed an approach that
uses \ac{TL} to pre-train \acp{CNN} with a large number of \ac{MD} images 
from simulations,
and
then uses a small number of \ac{TEM} images from experiments to fine-tune the \acp{CNN}.
This approach helps us overcome the challenge posed by this and other
experimental datasets that are too small or sparse to span across
the entire feature domain.
with sparse experimental dataset.
Specifically, we used a dataset of 20 unknown \ac{MD} images for testing 
the pre-trained model,
and another dataset of 20 unknown \ac{TEM} images for testing the fine-tuned models.
To ensure valid model accuracies
across the models,
the testing images 
were not seen by the models at the training, validation, or fine-tuning steps.

\subsection{Pre-train and fine-tune \acp{CNN}}

\begin{figure}
\centering
\includegraphics[clip=true,scale=0.25,width=1\linewidth]{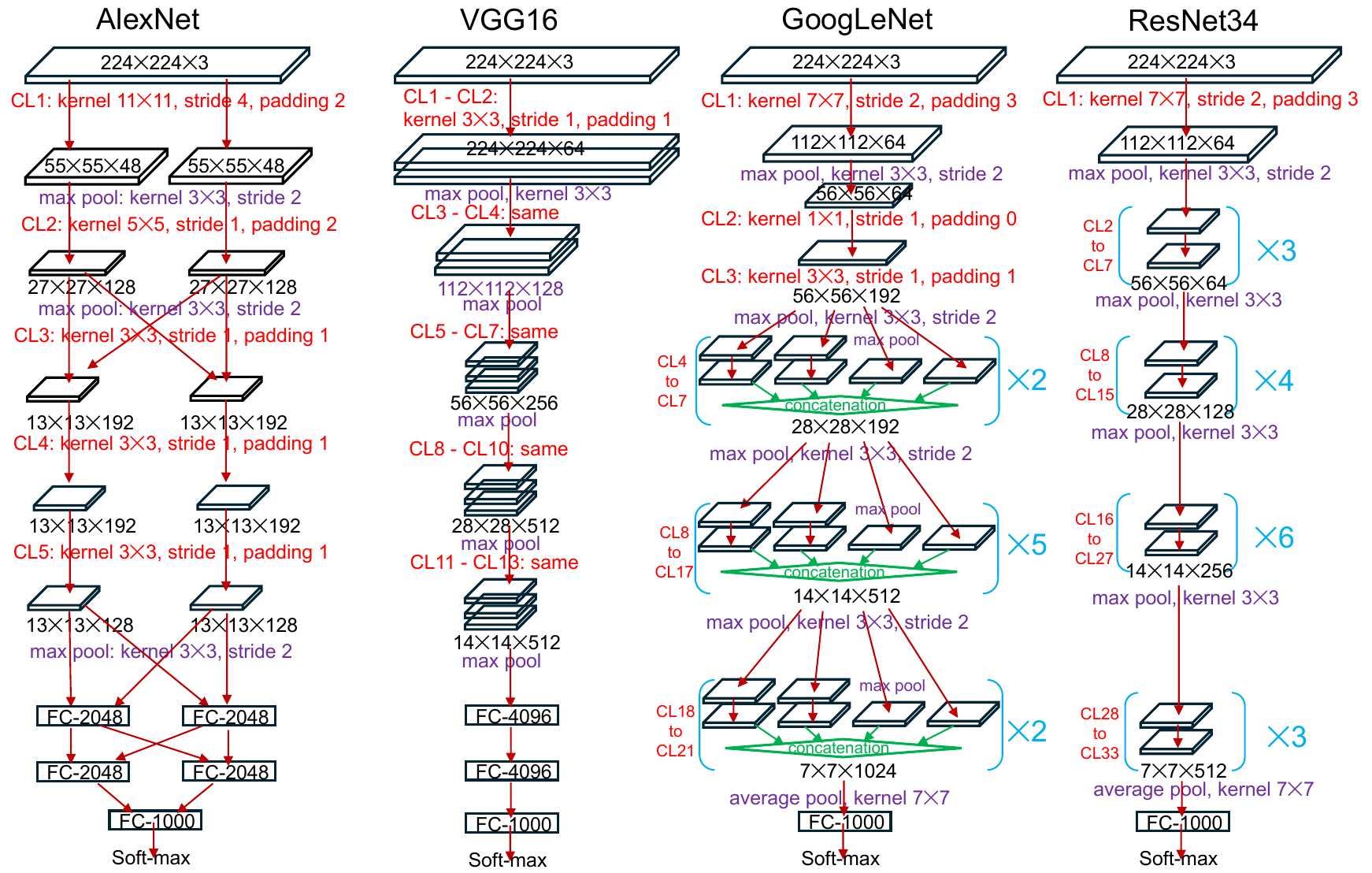}
\caption{
Comparison of the 4 benchmarking \acp{CNN} architectures for image classification:
AlexNet, VGG, GoogLeNet and ResNet.
VGG19 has 3 more \acp{CL} than VGG16.
ResNet18, ResNet50, ResNet101 and ResNet152 
vary the number of \acp{CL} with the same architecture as ResNet34.
The CL4 to CL21 in GoogLeNet use `kernel $1\times1$, stride 1 and padding 0',
`kernel $3\times3$, stride 1 and padding 1',
and `kernel $5\times5$, stride 1 and padding 2',
3 different \acp{CL}.
The CL2 to CL 33 in ResNet34 all use
`kernel $3\times3$, stride 1 and padding 1'.
The larger ResNets --- ResNet50, ResNet101 and ResNet152 --- use both
`kernel $1\times1$, stride 1 and padding 0' and
`kernel $3\times3$, stride 1 and padding 1' in \acp{CL}.
}
\label{fig:model}
\end{figure}

In this work, we implemented
the 4 different architectures---%
{\it viz.} AlexNet, VGG, GoogLeNet, and ResNet---%
shown in Figure~\ref{fig:model}.
We implement different sizes of the VGG and ResNet architectures---as noted
in the Figure caption---%
and report the performance of
9 different benchmarking \acp{CNN} (implemented in PyTorch\cite{pytorch})
in Figure \ref{fig:para}.
Such \ac{AI} models have precedent in 
nanomaterial applications;
\eg, Koyama et. al.\cite{koyama2021} used VGG16 to classify Pt-\acp{NP},
and Wang et. al.\cite{ywang2023} used ResNet50 to characterize DNA origamis.
Figure~\ref{fig:scheme} shows the scheme of the 
workflow for training \acp{AI} models.
First, 
720 images---that is, 120 for each of 
6 values of the ligation number---extracted
from \ac{MD} simulation trajectories were used to pre-train \acp{CNN}.
Then,
146 \ac{TEM} images---with 10-40 per ligation number---%
extracted from experiments 
were used to fine-tune the \acp{CNN}.
In the end, 
the fine-tuned \acp{CNN} were used to characterize unknown \ac{TEM} images. 
In the testing step,
the \ac{AI} model predictions were compared with human 
interpretation.

All images are resized to the same size of $224\times224\times3$---%
that is, across a $224\times224$ raster with each point
labeled by 3 color (RGB) intensities,---%
and processed by \acp{CNN} with different architectures;
see Figure~\ref{fig:model}.
At the pre-training step,
the \ac{MD} dataset was randomly split into $80\%$ training and $20\%$ validation images.
In both training and validation steps,
the datasets were loaded in batches of 32 images.
All models were run on one NVIDIA A100 40 GB GPU.
The total number of pre-training and fine-tuning epochs were both 90.
This was more than sufficient because
we found that by 40 epochs,
the training accuracy approaches 100\%,
the training loss approaches 0,
and the validation loss and accuracy approaches 100\%.
The cross entropy loss was used to pre-train and fine-tune \acp{CNN}.
The stochastic gradient decent optimizer was used in pre-training 
with a learning rate set at 0.001 
and a momentum set at 0.9.
The Adam optimizer was used in fine-tuning 
also with a learning rate set at 0.001.
All of the above hyper parameters were kept 
the same for different \acp{CNN}
in order to ensure equivalence in comparing across them.
We used 20 \ac{MD} and 20 \ac{TEM} unknown images
to calculate the testing accuracies 
of the pre-trained and fine-tuned \ac{CNN} models, respectively.

\section{Results and Discussion} 
\subsection{Architectures of \acp{CNN}}

\begin{figure}
\centering
\includegraphics[clip=true,scale=0.25,width=0.6\linewidth]{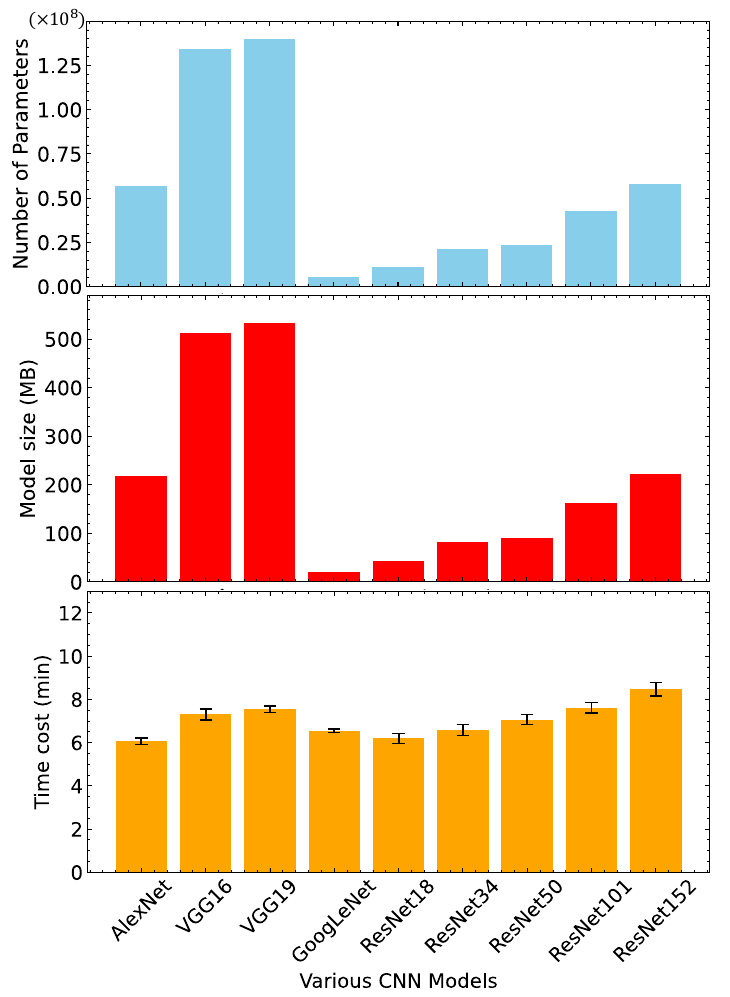}
\caption{Comparison of
total cost, model size and parameter size
across the 9 \acp{CNN}---{\it viz.}
AlexNet, VGG16, VGG19, GoogLeNet, ResNet18, ResNet34, ResNet50, 
ResNet101, and ResNet152---in
the bottom, middle and top panels, respectively.
The reported average computing time cost corresponds to the
completion of 90 epochs of training and validation runs 
on the \ac{MD} dataset using 1 NVIDIA A100 40 GB GPU.
}
\label{fig:para}
\end{figure}

The number of layers in
the present implementations of AlexNet, VGG16, GoogLeNet and ResNet34,
as summarized in Figure~\ref{fig:model},
are 5 \acp{CL} and 3 \acp{FC}, 
13 \acp{CL} and 3 \acp{FC}, 
21 \acp{CL} and 1 \ac{FC}, 
and 33 \acp{CL} and 1 \ac{FC}, respectively.
In the AlexNet architecture, 
the first \ac{CL} uses a kernel size of $11\times11$, stride 4 and padding 2;
the second \ac{CL} uses a kernel size of $5\times5$, stride 1 and padding 2;
and the third to fifth \acp{CL} use kernel size oof $3\times3$, stride 1 
and padding 1.
The VGG16 architecture has more layers than AlexNet,
and VGG19 has 3 more \acp{CL} than VGG16 with the same architecture.
As a result, the trend in the total number of parameters 
goes as:
$\mbox{VGG19} > \mbox{VGG16} > \mbox{AlexNet}$.
GoogLeNet
uses 9 repeated Inception Modules 
from the 4$^{\mbox{th}}$ to the 21$^{\mbox{st}}$ \acp{CL},
which significantly reduces the total number of parameters.
ResNet34 uses 4 repeated \ac{CL} blocks,
with the same kernel size ($3\times3$), stride 1 and padding 1 
from the 2$^{\mbox{nd}}$ to the 33$^{\mbox{rd}}$ \acp{CL}.
The ResNet architecture also significantly reduces
the total number of parameters.
Figure~\ref{fig:para} documents 
the increase in the total number of parameters 
from ResNet18 to ResNet152.
Notably, ResNet18 has lightly more parameters than GoogLeNet.
The model size is determined by the number of parameters
as can be verified by the similarity between them
across all of the \ac{CNN} models---%
as shown in the top and middle panels of
Figure~\ref{fig:para}.

All of the \ac{CNN} models were trained on NVIDIA A100 40 GB GPUs.
The computational time cost of pre-training and fine-tuning different \acp{CNN}
are reported in the bottom panel of Figure~\ref{fig:para}.
The performance comparison was conducted using one GPU card.
We found that even though the model sizes are significantly different,
the computational time costs are very close.
Specifically, the costs are about 6 to 9 min for finishing 90 training epochs 
and 90 fine-tuning epochs.
Notably, the size of GoogLeNet is 20 times less than VGG16 or VGG19,
but their time costs are within 15\% difference---%
{\it viz.} GoogLeNet is 6.5 min and VGG19 is 7.5 min.
The main reason is that 
deeper \ac{CNN} architectures---e.g., GoogLeNet and ResNet,---%
do not have much advantage
when used on research dataset
that are relatively small
even though they do perform well
on standard open datasets
which are necessarily large.
We also found that 
when the depth of \acp{CL} increases in ResNets,
the computational time costs increases monotonically.

\subsection{Training and validation accuracies}
 
\begin{figure}
\centering
\includegraphics[clip=true,scale=0.25,width=0.6\linewidth]{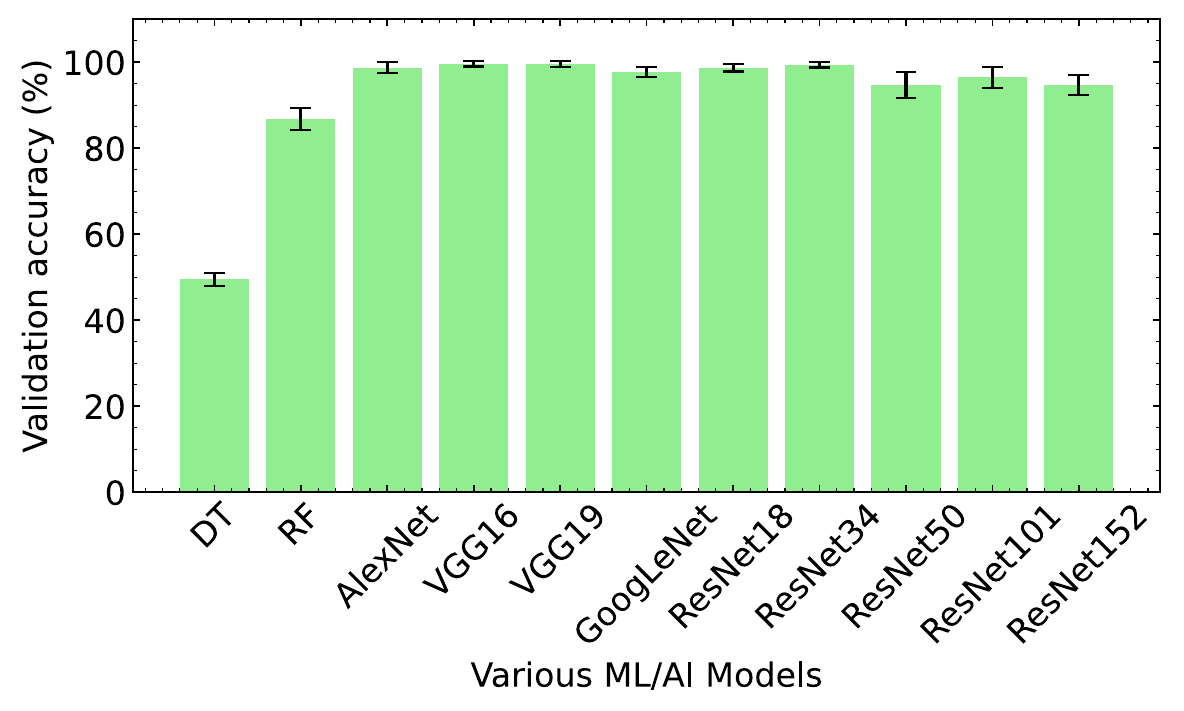}
\caption{
Validation accuracies among different \ac{ML} and \ac{AI} models---%
{\it viz.}
\ac{DT},\cite{song15} \ac{RF},\cite{breiman01} AlexNet, VGG16, VGG19, GoogLeNet, 
ResNet18, ResNet34, ResNet50, ResNet101, and ResNet152---%
after 90 training epochs.
}
\label{fig:val_acc}
\end{figure}

The training and validation accuracies at different epochs are 
available in Figure~S1 in \ac{SI}.
In all models, after 50 epochs,
the training accuracies converge to 100\%,
and validation accuracies also converge to some lower values.
The training and validation losses also shown  
in Figure~S1 in \ac{SI}
show the same trends.
The final validation accuracies at 90 epochs for different models 
are compared in Figure~\ref{fig:val_acc}.
We found that 
VGG16, VGG19 and ResNet34 have the best validation accuracies at $>99\%$;
AlexNet, GoogLeNet and ResNet18 have 97\% - 99\% validation accuracies;
and ResNet50, ResNet101 and ResNet152 have 94\% - 97\% validation accuracies.
In general, 
we show that all \acp{CNN} models have good validation accuracies $>94\%$, 
which is much higher than  that in the \ac{ML} models---%
{\it viz.} \ac{DT} is $<50\%$ and \ac{RF} is $<90\%$.
ResNet101 was earlier reported
to show higher accuracies than VGG16
across standard open datasets---%
e.g., PASCAL, COCO, and CIFAR-10.\cite{resnet15}
However, in our \ac{MD} image datasets,
VGG16 and VGG19 are slightly better than the ResNet architecture,
which may be due to 
the fact that VGG has more parameters; see Figure~\ref{fig:para}.
Thus, for a small research dataset,
we found that
it is not necessary to increase the depth of a \ac{CNN} 
so as to increase the accuracy
because 
all of the initial \ac{CNN} models 
started with sufficiently high validation accuracies from the outset.

\subsection{Testing accuracies on \ac{MD} images}

\begin{figure}
\centering
\includegraphics[clip=true,scale=0.25,width=1\linewidth]{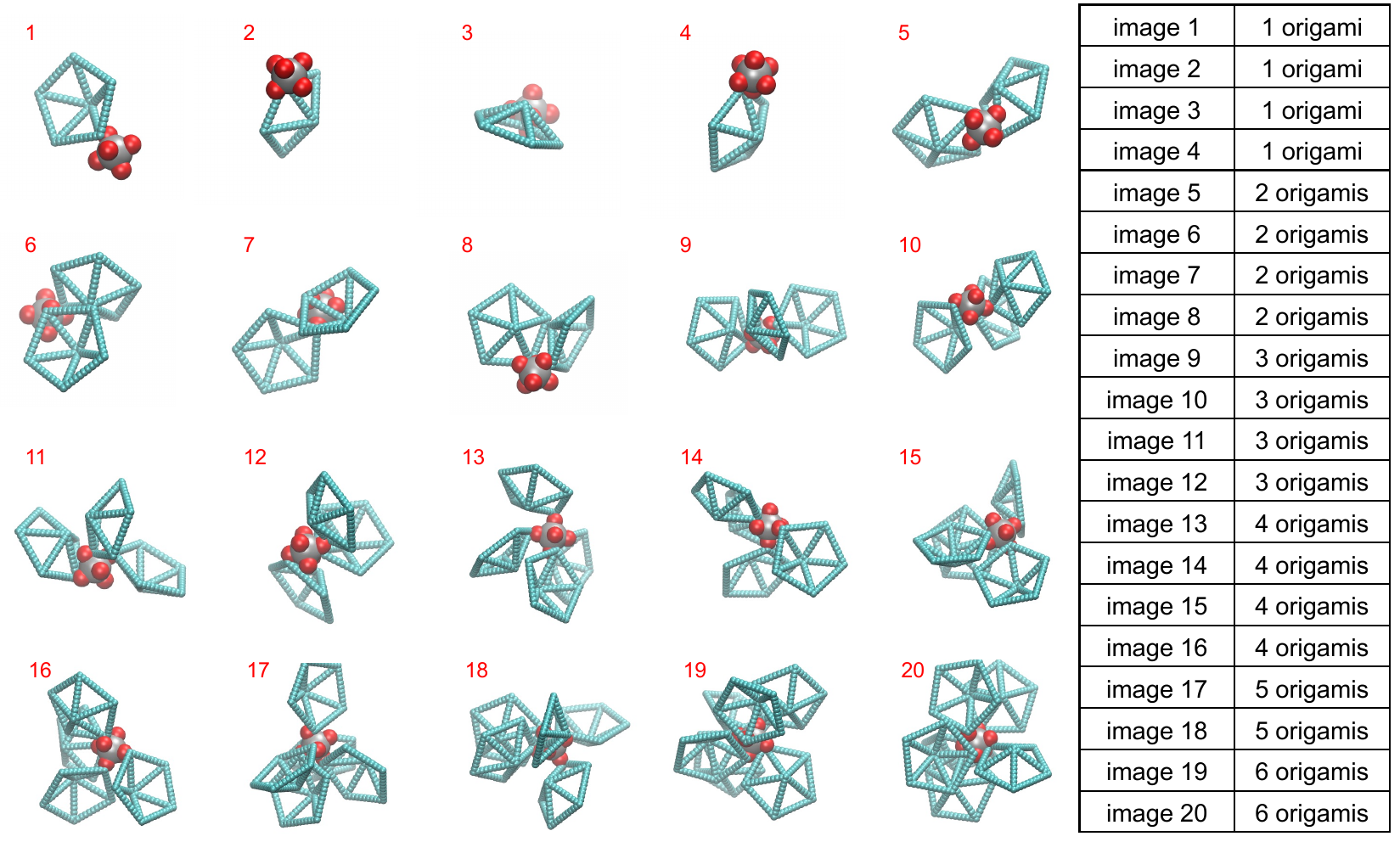}
\caption{%
The simulated test dataset consists of the
twenty images, 1-20, obtained
from projections of \ac{CG} \ac{MD} 3D models.
The ground truth for each image---%
{\it viz.} the ligation number---is listed
in the table on the right.
}
\label{fig:test1}
\end{figure}

\begin{figure}
\centering
\includegraphics[clip=true,scale=0.25,width=1\linewidth]{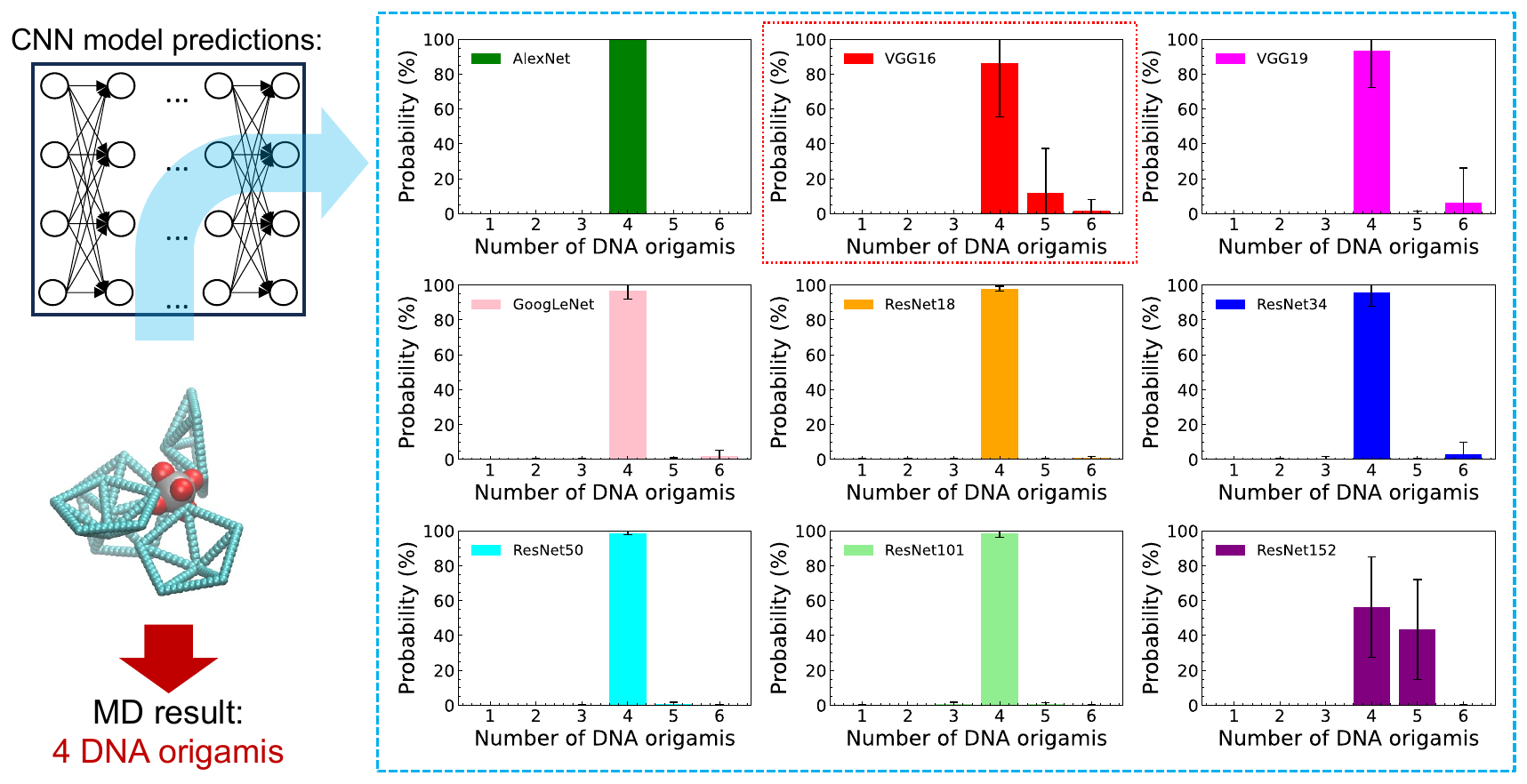}
\caption{
Comparison of the identification of an illustrative\ac{MD} trajectory image 
across 9 \acp{CNN}---{\it viz.}
AlexNet, VGG16, VGG19, GoogLeNet, ResNet18, ResNet34, ResNet50, 
ResNet101, and ResNet152.
The error bar is the standard deviation of 10 trainings using 
different random seeds.
}
\label{fig:classify1}
\end{figure}

\begin{figure}
\centering
\includegraphics[clip=true,scale=0.25,width=0.6\linewidth]{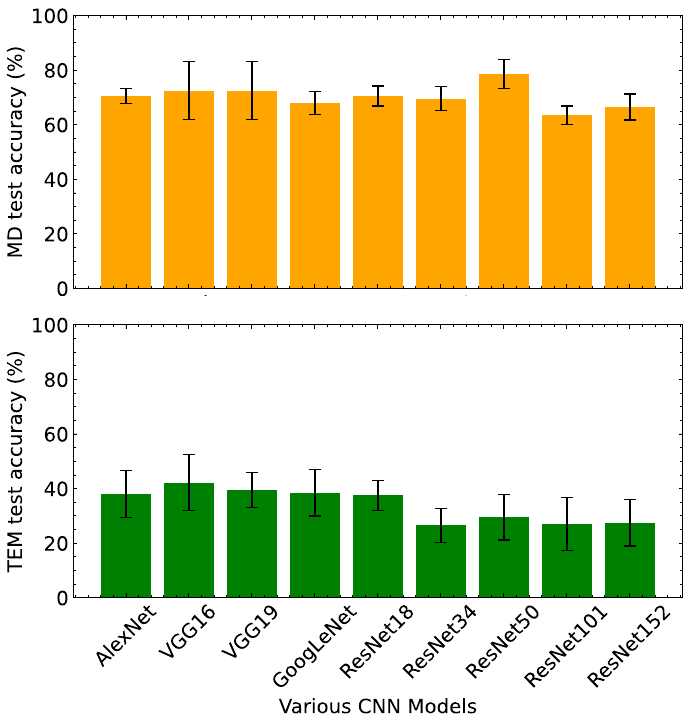}
\caption{
The accuracies across 9 \acp{CNN}
for identifying the test set of
20 \ac{MD} images shown in
Figure~\ref{fig:test1} (upper panel)
and the 20 \ac{TEM} images 
reported in Figure~\ref{fig:test2} (lower panel).
The error bar is the standard deviation of 10 trainings using 
different random seeds.
}
\label{fig:test_acc}
\end{figure}

Figure~\ref{fig:test1} shows the twenty MD images 
used for testing the pre-trained \acp{CNN}.
This set includes four replicates for 
for structures with 1 to 4 DNA origamis, 
and only two replicates for structures with 5 to 6 DNA origamis.
Test images are not seen by the models during training.
The ground truth for all \ac{MD} images are listed on the right in 
Figure~\ref{fig:test1}.
For example, 
nanostructure \#15 in Figure~\ref{fig:test1} has 4 DNA origamis 
bound to a \ac{QD}
as predicted correctly by all \ac{CNN} models 
are reported in Figure~\ref{fig:classify1}.
While most of the models correctly predict
that the structures contain 4 DNA origamis (at a probability of
90\% to 100\%) as reported in Figure~\ref{fig:classify1},
ResNet152 predicts that the structure
contains either 4 or 5 DNA origamis
with probabilities of 60\% and 40\%, respectively.
The testing accuracies 
across all twenty \ac{MD} images
for different pre-trained models are shown 
in the upper panel of Figure~\ref{fig:test_acc}.
The highest testing accuracy for pre-trained models 
is achieved by ResNet50 at $78.5\%$,
but the lowest testing accuracy is also achieved by ResNet101 at $63.5\%$.
The second best models are VGG16 and VGG19
as both have achieved $72.5\%$ accuracies.
Both AlexNet and ResNet18 also achieved $70.5\%$ accuracies.
This compares favorably with the testing accuracies seen
in pre-trained \ac{CNN} models of other
datasets.
For example, the highest top-1 prediction accuracies for 
AlexNet on the ILSVRC-2012 dataset was reported
at 63.3\%,\cite{alexnet12}
for VGG networks on the ILSVRC-2014 dataset
was reported at 75.3\%,\cite{vgg14}
and ResNet152 on the ILSVRC-2015 dataset
was reported at 80.62\%.\cite{resnet15}

\subsection{Testing accuracies on \ac{TEM} images}

\begin{figure}
\centering
\includegraphics[clip=true,scale=0.25,width=1\linewidth]{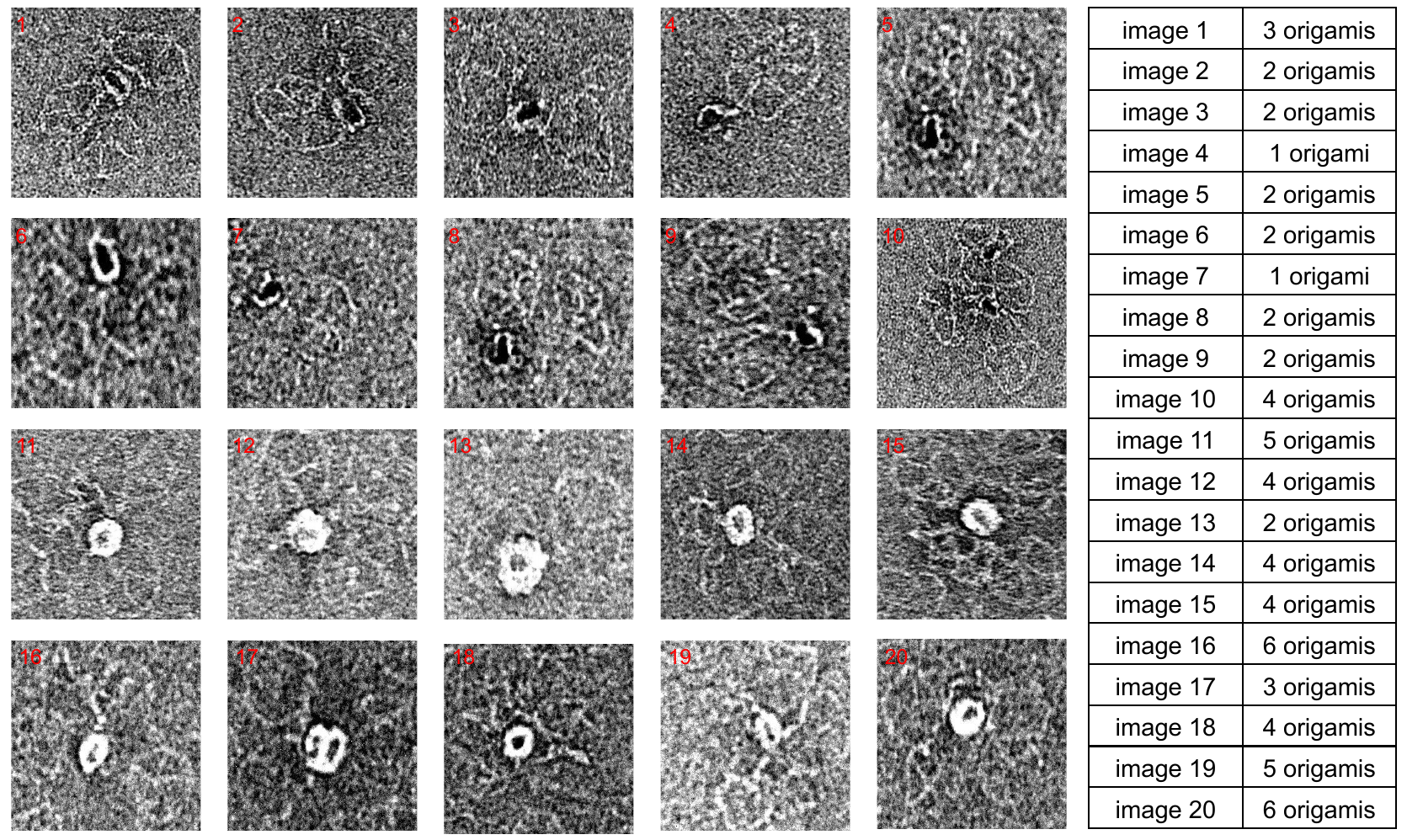}
\caption{%
The \ac{TEM} test dataset consists of the
twenty images, 1-20, obtained from measurement.
Scale bars are the same as shown in our previous paper.\cite{hern24g} 
The DNA origamis exhibit the same pentagonal pyramid shape 
as used in the \ac{CG} \ac{MD} models.
The true labels listed on the right are reported from observations.
}
\label{fig:test2}
\end{figure}

\begin{figure}
\centering
\includegraphics[clip=true,scale=0.25,width=1\linewidth]{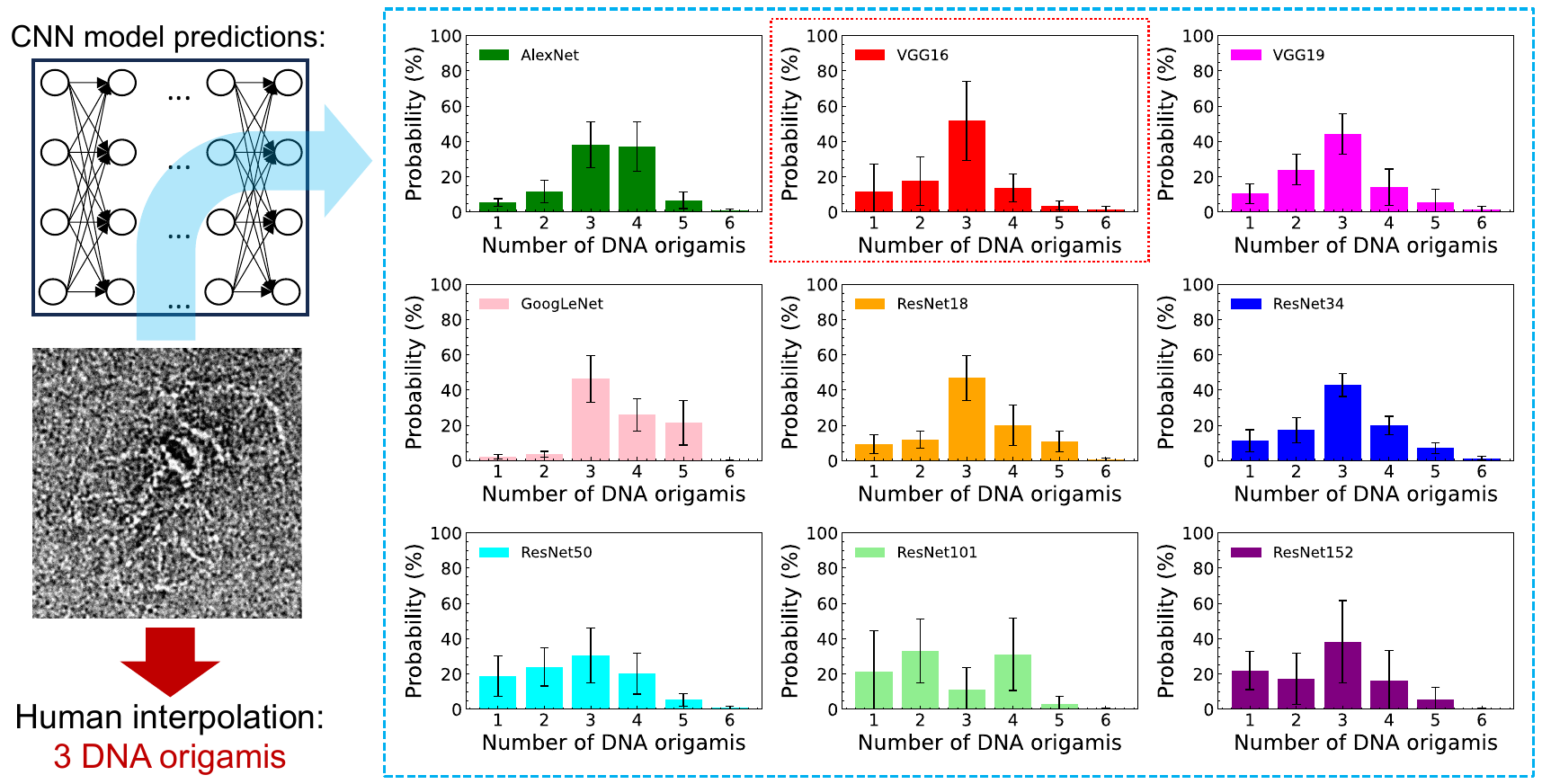}
\caption{%
A comparison of the identification of
a sample \ac{TEM} image using the 9 \acp{CNN}---{\it viz.}
AlexNet, VGG16, VGG19, GoogLeNet, ResNet18, ResNet34, ResNet50, 
ResNet101, and ResNet152.
The error bar is the standard deviation of 10 runs using different random seeds.
}
\label{fig:classify2}
\end{figure}

To characterize DNA origami nanostructures in experimental \ac{TEM} images
according to their ligation numbers,
pre-trained \ac{CNN} models are fine-tuned
using 146 \ac{TEM} images---whose ligation numbers have been assigned
by eye---%
with 10 to 40 images found ligation numbers between 1 and 6.
In the fine-tuning step,
all the layers are frozen
except for the last classifier layer 
whose unfrozen parameters are tuned.
This process allows us to take advantage of our simulation image dataset 
and transfer the useful parameters learned from \ac{MD} images
to train more reliable \ac{CNN} models using a small experimental dataset.
For example, in our \ac{TEM} images,
we found only 10 nanostructures images with 6 DNA origamis.
This is a consequence of the
very slow reaction rate from 5 DNA origamis to 6 DNA origamis 
when the \ac{SAv} binding site on the \ac{QD} is hindered. \cite{hern24g}

One challenge in identifying the ligation number of an experimental image
arises from the possible error in the identification of nanostructures 
by eye.
Especially, for images
containing a large number--- {\it viz.} between 4 to 6---of DNA origamis,
they are even harder to distinguish in the projected images;
see Figure~\ref{fig:scheme}.
As such, the observations are not quite the ground truth
and we are unable to establish absolute accuracy in
determining the ligation number on a given QD.
Instead, we report the relative agreement---and lack thereof---%
comparison between the observations and the \ac{ML} predictions.
In doing so, 
we demonstrate that 
the use of benchmarking \ac{AI} models 
can help accelerate
the identification of the ligation number
of DNA origami nanostructures in \ac{TEM} images.

As noted above, human bias is initially introduced
into the training data
when the l46 \ac{TEM} images are labeled.
Similarly, the 20 test images
are also characterized with the same human bias,
and those labels are listed on the right side of Figure~\ref{fig:test2}.
The uncertainty from the \ac{AI} model predictions 
for the testing data can be large 
because they are based on the 
training set which is itself biased, and possibly differently biased.
An advantage of the \ac{AI} models 
over assignments-by-eye is the fact that
the former can quickly determine the probability distribution
for the ligation number
for any given \ac{TEM} image.
For example, Figure~\ref{fig:classify2} shows that 
the \#1 nanostructure \ac{TEM} image is labelled as 3 DNA origamis 
by eye, and
that most of the AI models---namely,
VGG16, VGG19, GoogLeNet, ResNet18, ResNet34, ResNet50 and ResNet152---%
also identify it
as containing 3 DNA origamis with the highest probability;
see Figure~\ref{fig:classify2}.
On the other hand,
Figure~\ref{fig:classify2} also shows that 
AlexNet predicts that it has either 3 or 4 DNA origamis
with equal probabilities of $\sim40\%$,
and ResNet101 mistakenly predicts that it has either 2 or 4 DNA origamis.

We show the averaged testing 
agreements for the 20 \ac{TEM} images by different \acp{CNN}
in the bottom panel of Figure~\ref{fig:test_acc}:
VGG16 makes corresponding assignments
with the highest probability at 42.2\%, 
VGG19 is second at 39.4\%,
GoogLeNet is third at 38.5\%,
AlexNet is fourth at 38.0\%,
ResNet18 is fifth at 37.5\%,
and the other ResNets are all $<30\%$.
Because this agreement includes error
both in the training and testing sets, 
the agreement is not nearly as good 
as was found for the 
testing accuracy of the
\ac{MD}-based machines
reported in the upper panel of Figure~\ref{fig:test_acc}.
We may thus conclude that for small \ac{TEM} image datasets,
the best model with the highest relative (and possibly exact)
accuracy is the VGG architecture,
and it may result from having 
the largest model size.
We also found that
increasing the ResNet model depth does not 
improve the prediction accuracy.

\subsection{Classifying nanostructures in one large \ac{TEM} image}

\begin{figure}
\centering
\includegraphics[clip=true,scale=0.25,width=1\linewidth]{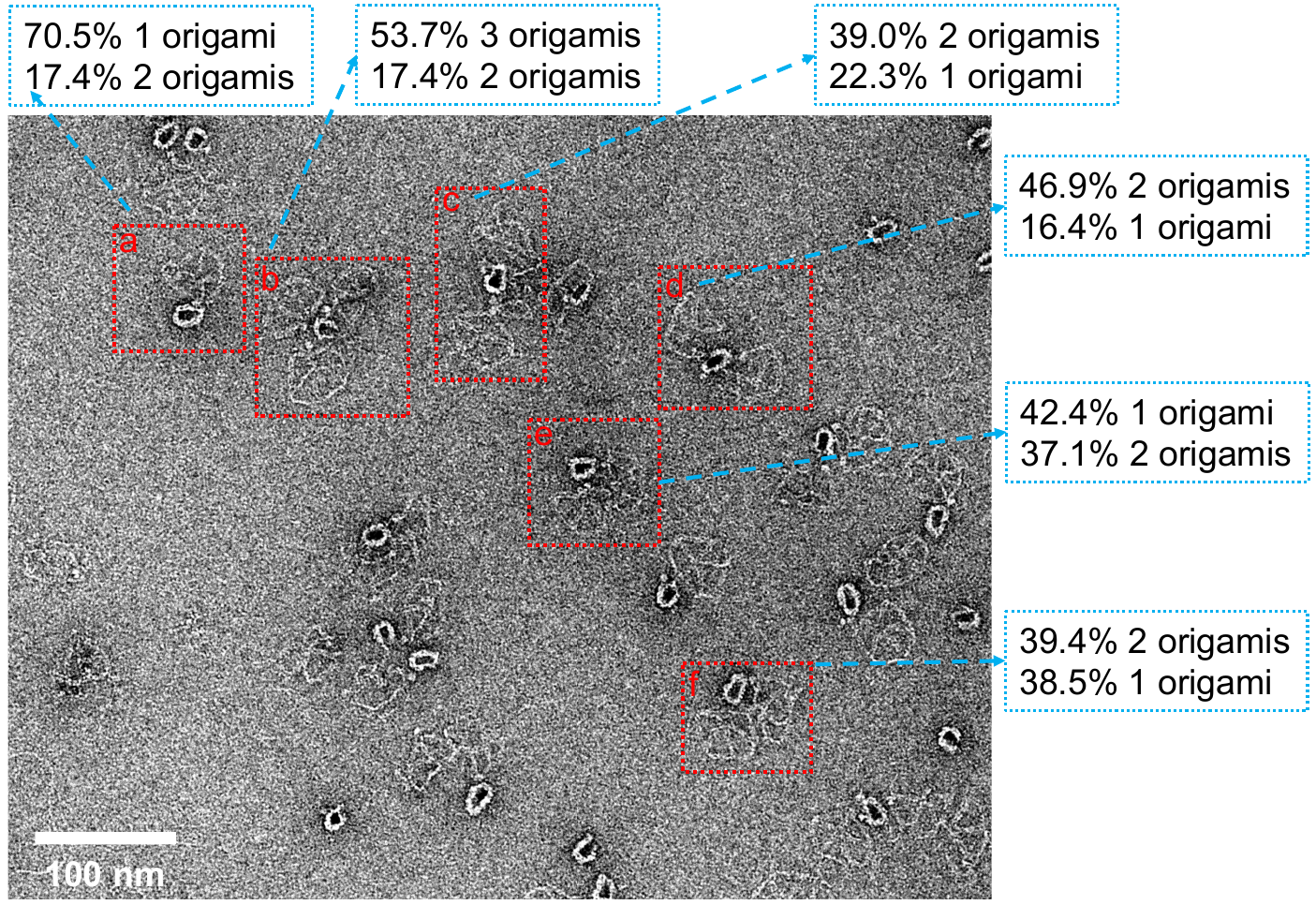}
\caption{%
Illustration of the characterizing of a typical, but large, TEM image using the VGG16 model.
Here, 6 different segments, labeled (a) - (f),
are identified from the big image for classification one-by-one.
In the figure, each structure is annotated by a blue box 
containing the two most likely assignments, 
noting the probability and the corresponding label.
These assignments are obtained from 
averages over 10 runs using different random seeds.
}
\label{fig:segment}
\end{figure}

Motivated by the success of VGG16 found in the previous section,
we now use it to characterize
multiple nanostructures in one large \ac{TEM} image,
and thereby confirm that it can resolve the identification of
the ligation number even in the presence of other nanostructures.
As an example, Figure~\ref{fig:segment} shows the 
manual selection of 
6 different regions, each containing a DNA origami nanostructure,
from a large \ac{TEM} image.
For simple nanostructures---%
{\it viz.} (a), (b), and (d) in Figure~\ref{fig:segment}---%
the fine-tuned VGG16 model can successfully identify their
ligation numbers,
with probability likelihoods of
70.5\%, 53.7\% and 46.9\% 
set to
2, 3 and 1 DNA origamis 
for regions (a), (b), and (d), respectively.
Meanwhile, in these three regions,
the second highest probability assignments
are much lower than the highest one,
and hence providing assurance of the identification.
For complex nanostructures---%
{\it viz.} (c), (e), and (f) in Figure~\ref{fig:segment}---%
the VGG16 model gives very close probabilities 
between the two most likely assignments,
predicting them with either 1 or 2 DNA origamis.
Thus, for more complex nanostructures---%
such as those in Figures~S2 - S4 in the \ac{SI}---%
we see that the trained VGG16 model can quickly 
reduce the number of possible assignments
for each segment, leaving the
eye to the simpler task of resolving between the reduced choices.
This suggests yet another
advantage of the \ac{AI} models in that 
they make predictions more consistent, faster, 
and with no or reduced human intervention or bias.

It should be noted
that YOLO models are commonly used for multiple object detection 
in large images
due to their efficiency in real-time detection. \cite{yolo}
For example, 
YOLOv5 was used for nanostructure detection in \ac{TEM} images.\cite{ywang2023}
However, when nanostructures are connected to each other
such as in 
the more complex \ac{TEM} images of
Figures~S2 - S4 in \ac{SI},
the use of YOLO models 
for performing segmentation and classification simultaneously
remains a big challenge.
It is also notable that 
the preparation of datasets
for training and fine-tuning a YOLO model for object detection 
is also much more time consuming
than the image classification models reported here.
Thus, for the moment, we favor the use of the ML and AI models
reported here for the classification of the ligation number
of a given nanostructure.

\FloatBarrier% Ensures that floats appear before conclusions.

\section{Conclusions}

In this work,
we demonstrate that 
supplementing 
a small dataset from experiments 
with 
a larger dataset from \ac{MD} simulations
can help train \ac{AI} models 
to characterize the ligation number of 
nanostructures in \ac{TEM} images.
We benchmark the performance of 9 \ac{CNN} models---{\it viz.}
AlexNet, GoogLeNet, VGG16, VGG19, ResNet18, 
ResNet34, ResNet50, ResNet101, and ResNet152.
We found that 
all \ac{CNN} models have similar computational time requirements,
even though their model sizes vary.
In the pre-training (and first) step the models are trained only
on \ac{MD} images.
All models reach $\sim100\%$ training accuracies and $>94\%$ validation accuracies
after 90 epochs with 720 \ac{MD} images.
Their testing accuracies are determined using 20 additional test \ac{MD} images.
We found that ResNet101, VGG16, and VGG19 have the top 3
testing accuracies of 78.5\%, 72.5\% and 72.5\%, respectively.
In the second and last step, we retrain the models by exposing
them to experimental structures whose ligation numbers was identified
by eye. 
Specifically,
we used 146 \ac{TEM} images from experiment to fine-tune the pre-trained models.
We also used 20 test \ac{TEM} images to determine the testing accuracies.
The exact ground truth for the 
ligation number identification of structures in the
\ac{TEM} images 
is unknown as they are made from visual inspection of the
2D projected images.
We thus resort to
reporting relative agreement between 
the model and test sets, and found that
the top 3 are VGG16, VGG19, and GoogLeNet at
42.2\%, 39.4\%, and 39.5\%, respectively.
We thus found that use of the fine-tuned VGG16 model
allows for quick
identification of the ligation numbers of nanostructures
in large \ac{TEM} images containing 
multiple nanostructures.

%%%%%%%%%%%%%%%%%%%%%%%%%%%%%%%%%%%%%%%%%%%%%%%%%%%%%%%

%\section*{Associated Content}

%\subsection*{Data Availability Statement}
%The code and datasets used for training, validation, and testing are available in
%the GitHub repository,
%\url{https://github.com/rxhernandez/TEMCOIN}.

%\subsection*{Supporting Information}
%The Supporting Information (SI) includes more information about methods, 
%and additional results to support the main text.

\begin{suppinfo}
The  \ac{SI} contains:
\begin{itemize}
\item More details about our computational methods.
(in Sec.~S-1 Additional Methods).
\item More results about characterization DNA origami nanostructures in large TEM images.
(in Sec.~S-2 Additional Results)
\end{itemize}
\end{suppinfo}
 
\section*{Data and Software Availability}
The code and datasets used for training, validation, and testing are available in
the GitHub repository,
\url{https://github.com/rxhernandez/TEMCOIN}.

\section*{Author Contributions}
\textbf{Dr. Xingfei Wei :}
Conceptualization,
Software Package Development,
Simulation Work, Dataset Preparation, 
Writing Original Draft, and Editing Paper.
\textbf{Qiankun Mo :} 
Conceptualization,
Software Package Development, Dataset Preparation, and Review Paper.
\textbf{Dr. Chi Chen :}
Conceptualization,
Experimental Work, Dataset Preparation, and Review Paper.
\textbf{Dr. Mark Bathe :}
Conceptualization,
Experimental Work, Review Paper, Supervision, and Funding Acquisition.
\textbf{Dr. Rigoberto Hernandez :}
Conceptualization,
Software Package Development, Simulation Work, Editing Paper, 
Supervision, and Funding Acquisition.

\section*{Competing interests}
The authors declare no competing financial interest.

\section*{Acknowledgments}
This work has been partially supported by the 
National Science Foundation (NSF) through Grant No.~CHE 2102455.
Funding support to MB and CC from ONR N00014-21-1-4013 
and NSF CCF 1956054 are also gratefully acknowledged.
The computing resources necessary for this work were
performed in part on Expanse 
at the San Diego Supercomputing Center
through allocation CTS090079 provided 
by \ac{ACCESS}, which is supported by National Science Foundation (NSF)
grants \#2138259, \#2138286, \#2138307, \#2137603, and \#2138296.
Additional computing resources
were provided by the Advanced Research Computing at Hopkins (ARCH) 
high-performance computing (HPC) facilities.

%\bibliography{j,japs,hern,acm,md}
\bibliography{DNA_origami_classification}

\providecommand{\latin}[1]{#1}
\makeatletter
\providecommand{\doi}
  {\begingroup\let\do\@makeother\dospecials
  \catcode`\{=1 \catcode`\}=2 \doi@aux}
\providecommand{\doi@aux}[1]{\endgroup\texttt{#1}}
\makeatother
\providecommand*\mcitethebibliography{\thebibliography}
\csname @ifundefined\endcsname{endmcitethebibliography}
  {\let\endmcitethebibliography\endthebibliography}{}
\begin{mcitethebibliography}{45}
\providecommand*\natexlab[1]{#1}
\providecommand*\mciteSetBstSublistMode[1]{}
\providecommand*\mciteSetBstMaxWidthForm[2]{}
\providecommand*\mciteBstWouldAddEndPuncttrue
  {\def\EndOfBibitem{\unskip.}}
\providecommand*\mciteBstWouldAddEndPunctfalse
  {\let\EndOfBibitem\relax}
\providecommand*\mciteSetBstMidEndSepPunct[3]{}
\providecommand*\mciteSetBstSublistLabelBeginEnd[3]{}
\providecommand*\EndOfBibitem{}
\mciteSetBstSublistMode{f}
\mciteSetBstMaxWidthForm{subitem}{(\alph{mcitesubitemcount})}
\mciteSetBstSublistLabelBeginEnd
  {\mcitemaxwidthsubitemform\space}
  {\relax}
  {\relax}

\bibitem[Hu \latin{et~al.}(2019)Hu, Li, Wang, Gu, and Fan]{CFan2019rev}
Hu,~Q.; Li,~H.; Wang,~L.; Gu,~H.; Fan,~C. DNA Nanotechnology-enabled Drug
  Delivery Systems. \emph{Chem. Rev.} \textbf{2019}, \emph{119}, 6459--6506,
  DOI: \doi{10.1021/acs.chemrev.7b00663}\relax
\mciteBstWouldAddEndPuncttrue
\mciteSetBstMidEndSepPunct{\mcitedefaultmidpunct}
{\mcitedefaultendpunct}{\mcitedefaultseppunct}\relax
\EndOfBibitem
\bibitem[Engelen and Dietz(2021)Engelen, and Dietz]{Engelen2021}
Engelen,~W.; Dietz,~H. Advancing Biophysics using DNA Origami. \emph{Annu. Rev.
  Biophys.} \textbf{2021}, \emph{50}, 469--492, DOI:
  \doi{10.1146/annurev-biophys-110520-125739}\relax
\mciteBstWouldAddEndPuncttrue
\mciteSetBstMidEndSepPunct{\mcitedefaultmidpunct}
{\mcitedefaultendpunct}{\mcitedefaultseppunct}\relax
\EndOfBibitem
\bibitem[Knappe \latin{et~al.}(2023)Knappe, Wamhoff, and Bathe]{Bathe2023}
Knappe,~G.~A.; Wamhoff,~E.-C.; Bathe,~M. Functionalizing DNA Origami to
  Investigate and Interact with Biological Systems. \emph{Nat. Rev. Mater.}
  \textbf{2023}, \emph{8}, 123--138, DOI:
  \doi{10.1038/s41578-022-00517-x}\relax
\mciteBstWouldAddEndPuncttrue
\mciteSetBstMidEndSepPunct{\mcitedefaultmidpunct}
{\mcitedefaultendpunct}{\mcitedefaultseppunct}\relax
\EndOfBibitem
\bibitem[Yao \latin{et~al.}(2015)Yao, Li, Chao, Pei, Liu, Zhao, Shi, Huang,
  Wang, Huang, and Fan]{CFan2015}
Yao,~G.; Li,~J.; Chao,~J.; Pei,~H.; Liu,~H.; Zhao,~Y.; Shi,~J.; Huang,~Q.;
  Wang,~L.; Huang,~W. \latin{et~al.}  Gold-Nanoparticle-Mediated
  Jigsaw-Puzzle-Like Assembly of Supersized Plasmonic DNA Origami. \emph{Angew.
  Chem., Int. Ed.} \textbf{2015}, \emph{54}, 2966--2969, DOI:
  \doi{10.1002/anie.201410895}\relax
\mciteBstWouldAddEndPuncttrue
\mciteSetBstMidEndSepPunct{\mcitedefaultmidpunct}
{\mcitedefaultendpunct}{\mcitedefaultseppunct}\relax
\EndOfBibitem
\bibitem[Bathe \latin{et~al.}(2021)Bathe, Hernandez, Komiyama, Machiraju, and
  Neogi]{hern21b}
Bathe,~M.; Hernandez,~R.; Komiyama,~T.; Machiraju,~R.; Neogi,~S. Autonomous
  Computing Materials. \emph{ACS Nano} \textbf{2021}, \emph{15}, 3586–3592,
  DOI: \doi{10.1021/acsnano.0c09556}\relax
\mciteBstWouldAddEndPuncttrue
\mciteSetBstMidEndSepPunct{\mcitedefaultmidpunct}
{\mcitedefaultendpunct}{\mcitedefaultseppunct}\relax
\EndOfBibitem
\bibitem[Dey \latin{et~al.}(2021)Dey, Fan, Gothelf, Li, Lin, Liu, Liu,
  Nijenhuis, Sacc{\`a}, Simmel, Yan, and Zhan]{CFan2021rev}
Dey,~S.; Fan,~C.; Gothelf,~K.~V.; Li,~J.; Lin,~C.; Liu,~L.; Liu,~N.;
  Nijenhuis,~M.~A.; Sacc{\`a},~B.; Simmel,~F.~C. \latin{et~al.}  DNA Origami.
  \emph{Nat. Rev. Methods Primers} \textbf{2021}, \emph{1}, 13, DOI:
  \doi{10.1038/s43586-020-00009-8}\relax
\mciteBstWouldAddEndPuncttrue
\mciteSetBstMidEndSepPunct{\mcitedefaultmidpunct}
{\mcitedefaultendpunct}{\mcitedefaultseppunct}\relax
\EndOfBibitem
\bibitem[Seeman(1982)]{seeman1982}
Seeman,~N.~C. Nucleic acid junctions and lattices. \emph{J. Theor. Biol.}
  \textbf{1982}, \emph{99}, 237--247, DOI:
  \doi{10.1016/0022-5193(82)90002-9}\relax
\mciteBstWouldAddEndPuncttrue
\mciteSetBstMidEndSepPunct{\mcitedefaultmidpunct}
{\mcitedefaultendpunct}{\mcitedefaultseppunct}\relax
\EndOfBibitem
\bibitem[Veneziano \latin{et~al.}(2016)Veneziano, Ratanalert, Zhang, Zhang,
  Yan, Chiu, and Bathe]{Bathe2016}
Veneziano,~R.; Ratanalert,~S.; Zhang,~K.; Zhang,~F.; Yan,~H.; Chiu,~W.;
  Bathe,~M. Designer Nanoscale DNA Assemblies Programmed from the Top Down.
  \emph{Science} \textbf{2016}, \emph{352}, 1534--1534, DOI:
  \doi{10.1126/science.aaf4388}\relax
\mciteBstWouldAddEndPuncttrue
\mciteSetBstMidEndSepPunct{\mcitedefaultmidpunct}
{\mcitedefaultendpunct}{\mcitedefaultseppunct}\relax
\EndOfBibitem
\bibitem[Bathe and Rothemund(2017)Bathe, and Rothemund]{Bathe2017}
Bathe,~M.; Rothemund,~P.~W. DNA Nanotechnology: A Foundation for Programmable
  Nanoscale Materials. \emph{MRS Bull.} \textbf{2017}, \emph{42}, 882--888,
  DOI: \doi{10.1557/mrs.2017.279}\relax
\mciteBstWouldAddEndPuncttrue
\mciteSetBstMidEndSepPunct{\mcitedefaultmidpunct}
{\mcitedefaultendpunct}{\mcitedefaultseppunct}\relax
\EndOfBibitem
\bibitem[Rothemund(2006)]{rothemund2006}
Rothemund,~P.~W. Folding DNA to Create Nanoscale Shapes and Patterns.
  \emph{Nature} \textbf{2006}, \emph{440}, 297--302, DOI:
  \doi{10.1038/nature04586}\relax
\mciteBstWouldAddEndPuncttrue
\mciteSetBstMidEndSepPunct{\mcitedefaultmidpunct}
{\mcitedefaultendpunct}{\mcitedefaultseppunct}\relax
\EndOfBibitem
\bibitem[Douglas \latin{et~al.}(2009)Douglas, Dietz, Liedl, H{\"o}gberg, Graf,
  and Shih]{douglas2009}
Douglas,~S.~M.; Dietz,~H.; Liedl,~T.; H{\"o}gberg,~B.; Graf,~F.; Shih,~W.~M.
  Self-Assembly of DNA into Nanoscale Three-Dimensional Shapes. \emph{Nature}
  \textbf{2009}, \emph{459}, 414--418, DOI: \doi{10.1038/nature08016}\relax
\mciteBstWouldAddEndPuncttrue
\mciteSetBstMidEndSepPunct{\mcitedefaultmidpunct}
{\mcitedefaultendpunct}{\mcitedefaultseppunct}\relax
\EndOfBibitem
\bibitem[Wang \latin{et~al.}(2022)Wang, Li, Jun, John, Zhang, Fowler, Doye,
  Chiu, and Bathe]{bathe2022wang}
Wang,~X.; Li,~S.; Jun,~H.; John,~T.; Zhang,~K.; Fowler,~H.; Doye,~J.~P.;
  Chiu,~W.; Bathe,~M. Planar {2D} Wireframe {DNA} Origami. \emph{Sci. Adv.}
  \textbf{2022}, \emph{8}, eabn0039, DOI: \doi{10.1126/sciadv.abn0039}\relax
\mciteBstWouldAddEndPuncttrue
\mciteSetBstMidEndSepPunct{\mcitedefaultmidpunct}
{\mcitedefaultendpunct}{\mcitedefaultseppunct}\relax
\EndOfBibitem
\bibitem[Chen \latin{et~al.}(2023)Chen, Luo, Kaplan, Bawendi, Macfarlane, and
  Bathe]{bathe2023cc}
Chen,~C.; Luo,~X.; Kaplan,~A.~E.; Bawendi,~M.~G.; Macfarlane,~R.~J.; Bathe,~M.
  Ultrafast Dense DNA Functionalization of Quantum Dots and Rods for Scalable
  2D Array Fabrication with Nanoscale Precision. \emph{Sci. Adv.}
  \textbf{2023}, \emph{9}, eadh8508, DOI: \doi{10.1126/sciadv.adh8508}\relax
\mciteBstWouldAddEndPuncttrue
\mciteSetBstMidEndSepPunct{\mcitedefaultmidpunct}
{\mcitedefaultendpunct}{\mcitedefaultseppunct}\relax
\EndOfBibitem
\bibitem[Chen \latin{et~al.}(2022)Chen, Wei, Parsons, Guo, Banal, Zhao, Scott,
  Schlau-Cohen, Hernandez, and Bathe]{hern22j}
Chen,~C.; Wei,~X.; Parsons,~M.~F.; Guo,~J.; Banal,~J.~L.; Zhao,~Y.;
  Scott,~M.~N.; Schlau-Cohen,~G.~S.; Hernandez,~R.; Bathe,~M. Nanoscale 3D
  Spatial Addressing and Valence Control of Quantum Dots Using Wireframe {DNA}
  Origami. \emph{Nat. Commun.} \textbf{2022}, \emph{13}, 4935, DOI:
  \doi{10.1038/s41467-022-32662-w}\relax
\mciteBstWouldAddEndPuncttrue
\mciteSetBstMidEndSepPunct{\mcitedefaultmidpunct}
{\mcitedefaultendpunct}{\mcitedefaultseppunct}\relax
\EndOfBibitem
\bibitem[Wei \latin{et~al.}(2024)Wei, Chen, Popov, Bathe, and
  Hernandez]{hern24g}
Wei,~X.; Chen,~C.; Popov,~A.; Bathe,~M.; Hernandez,~R. Binding Site
  Programmable Self-Assembly of {3D} Hierarchical {DNA} Origami Nanostructures.
  \emph{J. Phys. Chem. A} \textbf{2024}, \emph{128}, 4999–5008, DOI:
  \doi{10.1021/acs.jpca.4c02603}\relax
\mciteBstWouldAddEndPuncttrue
\mciteSetBstMidEndSepPunct{\mcitedefaultmidpunct}
{\mcitedefaultendpunct}{\mcitedefaultseppunct}\relax
\EndOfBibitem
\bibitem[Krizhevsky \latin{et~al.}(2017)Krizhevsky, Sutskever, and
  Hinton]{alexnet12}
Krizhevsky,~A.; Sutskever,~I.; Hinton,~G.~E. Imagenet Classification with Deep
  Convolutional Neural Networks. \emph{Commun. ACM} \textbf{2017}, \emph{60},
  84--90, DOI: \doi{10.1145/3065386}\relax
\mciteBstWouldAddEndPuncttrue
\mciteSetBstMidEndSepPunct{\mcitedefaultmidpunct}
{\mcitedefaultendpunct}{\mcitedefaultseppunct}\relax
\EndOfBibitem
\bibitem[Szegedy \latin{et~al.}(2015)Szegedy, Liu, Jia, Sermanet, Reed,
  Anguelov, Erhan, Vanhoucke, and Rabinovich]{googlenet15}
Szegedy,~C.; Liu,~W.; Jia,~Y.; Sermanet,~P.; Reed,~S.; Anguelov,~D.; Erhan,~D.;
  Vanhoucke,~V.; Rabinovich,~A. Going Deeper with Convolutions. 2015 IEEE
  Conference on Computer Vision and Pattern Recognition (CVPR). Los Alamitos,
  CA, USA, 2015; pp 1--9, DOI: \doi{10.1109/CVPR.2015.7298594}\relax
\mciteBstWouldAddEndPuncttrue
\mciteSetBstMidEndSepPunct{\mcitedefaultmidpunct}
{\mcitedefaultendpunct}{\mcitedefaultseppunct}\relax
\EndOfBibitem
\bibitem[Simonyan and Zisserman(2014)Simonyan, and Zisserman]{vgg14}
Simonyan,~K.; Zisserman,~A. Very Deep Convolutional Networks for Large-Scale
  Image Recognition. \emph{arXiv preprint arXiv:1409.1556} \textbf{2014}, DOI:
  \doi{10.48550/arXiv.1409.1556}\relax
\mciteBstWouldAddEndPuncttrue
\mciteSetBstMidEndSepPunct{\mcitedefaultmidpunct}
{\mcitedefaultendpunct}{\mcitedefaultseppunct}\relax
\EndOfBibitem
\bibitem[He \latin{et~al.}(2016)He, Zhang, Ren, and Sun]{resnet15}
He,~K.; Zhang,~X.; Ren,~S.; Sun,~J. Deep Residual Learning for Image
  Recognition. 2016 IEEE Conference on Computer Vision and Pattern Recognition
  (CVPR). Los Alamitos, CA, USA, 2016; pp 770--778, DOI:
  \doi{10.1109/CVPR.2016.90}\relax
\mciteBstWouldAddEndPuncttrue
\mciteSetBstMidEndSepPunct{\mcitedefaultmidpunct}
{\mcitedefaultendpunct}{\mcitedefaultseppunct}\relax
\EndOfBibitem
\bibitem[Russakovsky \latin{et~al.}(2015)Russakovsky, Deng, Su, Krause,
  Satheesh, Ma, Huang, Karpathy, Khosla, Bernstein, \latin{et~al.}
  others]{russakovsky15}
Russakovsky,~O.; Deng,~J.; Su,~H.; Krause,~J.; Satheesh,~S.; Ma,~S.; Huang,~Z.;
  Karpathy,~A.; Khosla,~A.; Bernstein,~M. \latin{et~al.}  ImageNet Large Scale
  Visual Recognition Challenge. \emph{Int. J. Comput. Vis.} \textbf{2015},
  \emph{115}, 211--252, DOI: \doi{10.1007/s11263-015-0816-y}\relax
\mciteBstWouldAddEndPuncttrue
\mciteSetBstMidEndSepPunct{\mcitedefaultmidpunct}
{\mcitedefaultendpunct}{\mcitedefaultseppunct}\relax
\EndOfBibitem
\bibitem[Li \latin{et~al.}(2022)Li, Andreeto, Ranzato, and Perona]{caltech101}
Li,~F.-F.; Andreeto,~M.; Ranzato,~M.; Perona,~P. Caltech 101 (1.0) [Data set].
  \emph{CaltechDATA} \textbf{2022}, DOI: \doi{10.22002/D1.20086}\relax
\mciteBstWouldAddEndPuncttrue
\mciteSetBstMidEndSepPunct{\mcitedefaultmidpunct}
{\mcitedefaultendpunct}{\mcitedefaultseppunct}\relax
\EndOfBibitem
\bibitem[Griffin \latin{et~al.}(2022)Griffin, Holub, and Perona]{caltech256}
Griffin,~G.; Holub,~A.; Perona,~P. Caltech 256 (1.0) [Data set].
  \emph{CaltechDATA} \textbf{2022}, DOI: \doi{10.22002/D1.20087}\relax
\mciteBstWouldAddEndPuncttrue
\mciteSetBstMidEndSepPunct{\mcitedefaultmidpunct}
{\mcitedefaultendpunct}{\mcitedefaultseppunct}\relax
\EndOfBibitem
\bibitem[Krizhevsky(2009)]{cifar10}
Krizhevsky,~A. Learning Multiple Layers of Features from Tiny Images. M.Sc.\
  thesis, Department of Computer Science, University of Toronto, 2009\relax
\mciteBstWouldAddEndPuncttrue
\mciteSetBstMidEndSepPunct{\mcitedefaultmidpunct}
{\mcitedefaultendpunct}{\mcitedefaultseppunct}\relax
\EndOfBibitem
\bibitem[Lee \latin{et~al.}(2021)Lee, Jeong, and Yang]{jlee2021}
Lee,~J.; Jeong,~C.; Yang,~Y. Single-Atom Level Determination of 3-Dimensional
  Surface Atomic Atructure via Neural Network-Assisted Atomic Electron
  Tomography. \emph{Nat. Commun.} \textbf{2021}, \emph{12}, 1962, DOI:
  \doi{10.1038/s41467-021-22204-1}\relax
\mciteBstWouldAddEndPuncttrue
\mciteSetBstMidEndSepPunct{\mcitedefaultmidpunct}
{\mcitedefaultendpunct}{\mcitedefaultseppunct}\relax
\EndOfBibitem
\bibitem[Nikishin \latin{et~al.}(2021)Nikishin, Dulimov, Skryabin, Galetsky,
  Tchevkina, and Bagrov]{nikishin2021}
Nikishin,~I.; Dulimov,~R.; Skryabin,~G.; Galetsky,~S.; Tchevkina,~E.;
  Bagrov,~D. ScanEV – A Neural Network-Based Tool for The Automated Detection
  of Extracellular Eesicles in TEM Images. \emph{Micron} \textbf{2021},
  \emph{145}, 103044, DOI: \doi{10.1016/j.micron.2021.103044}\relax
\mciteBstWouldAddEndPuncttrue
\mciteSetBstMidEndSepPunct{\mcitedefaultmidpunct}
{\mcitedefaultendpunct}{\mcitedefaultseppunct}\relax
\EndOfBibitem
\bibitem[Koyama \latin{et~al.}(2021)Koyama, Miyauchi, Morooka, Hojo, Einaga,
  and Murakami]{koyama2021}
Koyama,~A.; Miyauchi,~S.; Morooka,~K.; Hojo,~H.; Einaga,~H.; Murakami,~Y.
  Analysis of TEM Images of Metallic Nanoparticles Using Convolutional Neural
  Networks and Transfer Learning. \emph{J. Magn. Magn. Mater.} \textbf{2021},
  \emph{538}, 168225, DOI: \doi{10.1016/j.jmmm.2021.168225}\relax
\mciteBstWouldAddEndPuncttrue
\mciteSetBstMidEndSepPunct{\mcitedefaultmidpunct}
{\mcitedefaultendpunct}{\mcitedefaultseppunct}\relax
\EndOfBibitem
\bibitem[Sytwu \latin{et~al.}(2022)Sytwu, Groschner, and Scott]{sytwu2022}
Sytwu,~K.; Groschner,~C.; Scott,~M.~C. Understanding The Influence of Receptive
  Field and Network Complexity in Neural Network-Guided TEM Image Analysis.
  \emph{Microsc Microanal.} \textbf{2022}, \emph{28}, 1896--1904, DOI:
  \doi{10.1017/S1431927622012466}\relax
\mciteBstWouldAddEndPuncttrue
\mciteSetBstMidEndSepPunct{\mcitedefaultmidpunct}
{\mcitedefaultendpunct}{\mcitedefaultseppunct}\relax
\EndOfBibitem
\bibitem[Wang \latin{et~al.}(2023)Wang, Jin, and Castro]{ywang2023}
Wang,~Y.; Jin,~X.; Castro,~C. Accelerating The Characterization of Dynamic DNA
  Origami Devices with Deep Neural Networks. \emph{Sci. Rep.} \textbf{2023},
  \emph{13}, 15196, DOI: \doi{10.1038/s41598-023-41459-w}\relax
\mciteBstWouldAddEndPuncttrue
\mciteSetBstMidEndSepPunct{\mcitedefaultmidpunct}
{\mcitedefaultendpunct}{\mcitedefaultseppunct}\relax
\EndOfBibitem
\bibitem[Liu \latin{et~al.}(2023)Liu, Chen, Zhang, Zhang, Yang, Hu, Xiao, Liu,
  and Jiang]{lliu2023}
Liu,~L.; Chen,~T.; Zhang,~Q.; Zhang,~W.; Yang,~H.; Hu,~X.; Xiao,~J.; Liu,~Q.;
  Jiang,~G. Deep Neural Network-Based Electron Microscopy Image Recognition for
  Source Distinguishing of Anthropogenic and Natural Magnetic Particles.
  \emph{Environ. Sci. Technol.} \textbf{2023}, \emph{57}, 16465--16476, DOI:
  \doi{10.1021/acs.est.3c05252}\relax
\mciteBstWouldAddEndPuncttrue
\mciteSetBstMidEndSepPunct{\mcitedefaultmidpunct}
{\mcitedefaultendpunct}{\mcitedefaultseppunct}\relax
\EndOfBibitem
\bibitem[Gumbiowski \latin{et~al.}(2024)Gumbiowski, Barthel, Loza, Heggen, and
  Epple]{gumbiowski2024}
Gumbiowski,~N.; Barthel,~J.; Loza,~K.; Heggen,~M.; Epple,~M. Simulated HRTEM
  Images of Nanoparticles to Train A Neural Network to Classify Nanoparticles
  for Crystallinity. \emph{Nanoscale Adv.} \textbf{2024}, \emph{6}, 4196--4206,
  DOI: \doi{10.1039/d4na00266k}\relax
\mciteBstWouldAddEndPuncttrue
\mciteSetBstMidEndSepPunct{\mcitedefaultmidpunct}
{\mcitedefaultendpunct}{\mcitedefaultseppunct}\relax
\EndOfBibitem
\bibitem[Senanayake \latin{et~al.}(2022)Senanayake, Yao, Froehlich, Cahill,
  Sheldon, McIntire, Haynes, and Hernandez]{hern22m}
Senanayake,~R.~D.; Yao,~X.; Froehlich,~C.; Cahill,~M.~S.; Sheldon,~T.;
  McIntire,~M.; Haynes,~C.~L.; Hernandez,~R. Machine Learning-Assisted Carbon
  Dot Synthesis: {P}rediction of Emission Color and Wavelength. \emph{J. Chem.
  Inf. Model.} \textbf{2022}, \emph{62}, 5918–5928, DOI:
  \doi{10.1021/acs.jcim.2c01007}\relax
\mciteBstWouldAddEndPuncttrue
\mciteSetBstMidEndSepPunct{\mcitedefaultmidpunct}
{\mcitedefaultendpunct}{\mcitedefaultseppunct}\relax
\EndOfBibitem
\bibitem[Senanayake \latin{et~al.}(2024)Senanayake, {Daly, Jr.}, and
  Hernandez]{hern24e}
Senanayake,~R.~D.; {Daly, Jr.},~C.~A.; Hernandez,~R. Optimized Bags of
  Artificial Neural Networks Can Predict Viability of Organisms Exposed to
  Nanoparticles. \emph{J. Phys. Chem. A} \textbf{2024}, \emph{128}, 2857--2870,
  DOI: \doi{10.1021/acs.jpca.3c07462}\relax
\mciteBstWouldAddEndPuncttrue
\mciteSetBstMidEndSepPunct{\mcitedefaultmidpunct}
{\mcitedefaultendpunct}{\mcitedefaultseppunct}\relax
\EndOfBibitem
\bibitem[Thota \latin{et~al.}(2024)Thota, Priyadarshini, and
  Hernandez]{hern24b}
Thota,~N.~K.; Priyadarshini,~M.~S.; Hernandez,~R. {NestedAE}: Interpretable
  Nested Autoencoders for Multi-Scale Material Characterization. \emph{Mater.
  Horiz.} \textbf{2024}, \emph{11}, 700--707, DOI:
  \doi{10.1039/D3MH01484C}\relax
\mciteBstWouldAddEndPuncttrue
\mciteSetBstMidEndSepPunct{\mcitedefaultmidpunct}
{\mcitedefaultendpunct}{\mcitedefaultseppunct}\relax
\EndOfBibitem
\bibitem[Priyadarshini \latin{et~al.}(2025)Priyadarshini, Thota, and
  Hernandez]{hern25a}
Priyadarshini,~M.~S.; Thota,~N.~K.; Hernandez,~R. {ReLMM}:Reinforcement
  Learning Optimizes Feature Selection in Modeling Materials. \emph{J. Chem.
  Inf. Model.} \textbf{2025}, \emph{65}, 153--161, DOI:
  \doi{10.1021/acs.jcim.4c01934}\relax
\mciteBstWouldAddEndPuncttrue
\mciteSetBstMidEndSepPunct{\mcitedefaultmidpunct}
{\mcitedefaultendpunct}{\mcitedefaultseppunct}\relax
\EndOfBibitem
\bibitem[Wei \latin{et~al.}(2022)Wei, Chen, Zhao, Harazinska, Bathe, and
  Hernandez]{hern22e}
Wei,~X.; Chen,~C.; Zhao,~Y.; Harazinska,~E.; Bathe,~M.; Hernandez,~R. Molecular
  Structure of Single-Stranded {DNA} on the {ZnS} Surface of Quantum Dots.
  \emph{ACS Nano} \textbf{2022}, \emph{16}, 6666–6675, DOI:
  \doi{10.1021/acsnano.2c01178}\relax
\mciteBstWouldAddEndPuncttrue
\mciteSetBstMidEndSepPunct{\mcitedefaultmidpunct}
{\mcitedefaultendpunct}{\mcitedefaultseppunct}\relax
\EndOfBibitem
\bibitem[Plimpton(1995)]{plimpton95}
Plimpton,~S.~J. Fast Parallel Algorithms for Short-Range Molecular Dynamics.
  \emph{J. Comput. Phys.} \textbf{1995}, \emph{117}, 1--19, DOI:
  \doi{10.1006/jcph.1995.1039}\relax
\mciteBstWouldAddEndPuncttrue
\mciteSetBstMidEndSepPunct{\mcitedefaultmidpunct}
{\mcitedefaultendpunct}{\mcitedefaultseppunct}\relax
\EndOfBibitem
\bibitem[Weeks \latin{et~al.}(1971)Weeks, Chandler, and Andersen]{wca1971}
Weeks,~J.~D.; Chandler,~D.; Andersen,~H.~C. Role of Repulsive Forces in
  Determining the Equilibrium Structure of Simple Liquids. \emph{J. Chem.
  Phys.} \textbf{1971}, \emph{54}, 5237--5247, DOI:
  \doi{10.1063/1.1674820}\relax
\mciteBstWouldAddEndPuncttrue
\mciteSetBstMidEndSepPunct{\mcitedefaultmidpunct}
{\mcitedefaultendpunct}{\mcitedefaultseppunct}\relax
\EndOfBibitem
\bibitem[Humphrey \latin{et~al.}(1996)Humphrey, Dalke, and Schulten]{vmd}
Humphrey,~W.; Dalke,~A.; Schulten,~K. {VMD} --- {V}isual {M}olecular
  {D}ynamics. 1996; \url{https://www.ks.uiuc.edu/Research/vmd/}\relax
\mciteBstWouldAddEndPuncttrue
\mciteSetBstMidEndSepPunct{\mcitedefaultmidpunct}
{\mcitedefaultendpunct}{\mcitedefaultseppunct}\relax
\EndOfBibitem
\bibitem[Shepherd \latin{et~al.}(2019)Shepherd, Du, Huang, Wamhoff, and
  Bathe]{shepherd2019}
Shepherd,~T.~R.; Du,~R.~R.; Huang,~H.; Wamhoff,~E.-C.; Bathe,~M. Bioproduction
  of Pure, Kilobase-Scale single-stranded DNA. \emph{Sci. Rep.} \textbf{2019},
  \emph{9}, 6121, DOI: \doi{10.1038/s41598-019-42665-1}\relax
\mciteBstWouldAddEndPuncttrue
\mciteSetBstMidEndSepPunct{\mcitedefaultmidpunct}
{\mcitedefaultendpunct}{\mcitedefaultseppunct}\relax
\EndOfBibitem
\bibitem[{The GIMP Development Team}()]{gimp}
{The GIMP Development Team}, GIMP. \url{https://www.gimp.org}\relax
\mciteBstWouldAddEndPuncttrue
\mciteSetBstMidEndSepPunct{\mcitedefaultmidpunct}
{\mcitedefaultendpunct}{\mcitedefaultseppunct}\relax
\EndOfBibitem
\bibitem[Paszke \latin{et~al.}(2019)Paszke, Gross, Massa, Lerer, Bradbury,
  Chanan, Killeen, Lin, Gimelshein, Antiga, Desmaison, Kopf, Yang, DeVito,
  Raison, Tejani, Chilamkurthy, Steiner, Fang, Bai, and Chintala]{pytorch}
Paszke,~A.; Gross,~S.; Massa,~F.; Lerer,~A.; Bradbury,~J.; Chanan,~G.;
  Killeen,~T.; Lin,~Z.; Gimelshein,~N.; Antiga,~L. \latin{et~al.}  PyTorch: An
  Imperative Style, High-Performance Deep Learning Library. Advances in Neural
  Information Processing Systems 32. 2019; pp 8024--8035\relax
\mciteBstWouldAddEndPuncttrue
\mciteSetBstMidEndSepPunct{\mcitedefaultmidpunct}
{\mcitedefaultendpunct}{\mcitedefaultseppunct}\relax
\EndOfBibitem
\bibitem[Song and Ying(2015)Song, and Ying]{song15}
Song,~Y.-Y.; Ying,~L. Decision Tree Methods: Applications for Classification
  and Prediction. \emph{Shanghai Arch. Psychiatry} \textbf{2015}, \emph{27},
  130, DOI: \doi{10.11919/j.issn.1002-0829.215044}\relax
\mciteBstWouldAddEndPuncttrue
\mciteSetBstMidEndSepPunct{\mcitedefaultmidpunct}
{\mcitedefaultendpunct}{\mcitedefaultseppunct}\relax
\EndOfBibitem
\bibitem[Breiman(2001)]{breiman01}
Breiman,~L. Random Forests. \emph{Mach. Learn.} \textbf{2001}, \emph{45},
  5--32, DOI: \doi{10.1023/A:1010933404324}\relax
\mciteBstWouldAddEndPuncttrue
\mciteSetBstMidEndSepPunct{\mcitedefaultmidpunct}
{\mcitedefaultendpunct}{\mcitedefaultseppunct}\relax
\EndOfBibitem
\bibitem[Redmon \latin{et~al.}(2016)Redmon, Divvala, Girshick, and
  Farhadi]{yolo}
Redmon,~J.; Divvala,~S.; Girshick,~R.; Farhadi,~A. You Only Look Once: Unified,
  Real-Time Object Detection. 2016 IEEE Conference on Computer Vision and
  Pattern Recognition (CVPR). Los Alamitos, CA, USA, 2016; pp 779--788, DOI:
  \doi{10.1109/CVPR.2016.91}\relax
\mciteBstWouldAddEndPuncttrue
\mciteSetBstMidEndSepPunct{\mcitedefaultmidpunct}
{\mcitedefaultendpunct}{\mcitedefaultseppunct}\relax
\EndOfBibitem
\end{mcitethebibliography}

\end{document}